%% file: main.tex
\documentclass[sigconf, nonacm]{acmart}

\usepackage{blindtext}
\usepackage[breakable]{tcolorbox}

\usepackage{graphicx}
\usepackage{subcaption}
\usepackage{mathtools}
\usepackage{amsmath}
\usepackage{pdfpages}
\usepackage{svg}
\usepackage{enumitem}
\usepackage{tikz}
\usepackage{framed}
\usepackage{fancybox}
\usepackage{multirow}
\usepackage{tabularx}
\usepackage{hyperref}
\usepackage{xurl}

\usepackage{zref}
\usepackage{cleveref}
\crefformat{section}{\S#2#1#3}
\AtBeginDocument{%
  \providecommand\BibTeX{{%
    \normalfont B\kern-0.5em{\scshape i\kern-0.25em b}\kern-0.8em\TeX}}}

\setcopyright{acmcopyright}
\copyrightyear{2018}
\acmYear{2018}
\acmDOI{XXXXXXX.XXXXXXX}

\acmConference[Conference acronym 'XX]{Make sure to enter the correct
  conference title from your rights confirmation email}{June 03--05,
  2018}{Woodstock, NY}

\acmPrice{15.00}
\acmISBN{978-1-4503-XXXX-X/18/06}

\usepackage{xcolor}
\newcommand*\circled[1]{\tikz[baseline=(char.base)]{
            \node[shape=circle,fill,inner sep=1pt] (char) {\textcolor{white}{#1}};}}

\usepackage{xspace}
\newcommand{\alg}{\textsc{Elf}\xspace} 
\newcommand{\wf}{\textsc{Elves}\xspace}  

\newcommand{\btr}{\textsc{B\footnotesize{TR}}\textsc{B\footnotesize{LOCKS}}\xspace}

\newcounter{greyboxrq}

\begin{document}

{\title{Everything You Always Wanted to Know About Storage Compressibility of Pre-Trained ML Models but Were Afraid to Ask}}

\author{Zhaoyuan Su$^1$, ~~~ Ammar Ahmed$^2$, ~~~  Zirui Wang$^1$, ~~~ Ali Anwar$^2$, ~~~ Yue Cheng$^1$}
\affiliation{
    \institution{$^1${\it University of Virginia}, $^2${\it University of Minnesota}} 
}

\begin{abstract}
As the number of pre-trained machine learning (ML) models is growing exponentially, data reduction tools are not catching up.  
Existing data reduction techniques are not specifically designed for pre-trained model {(PTM) dataset files}. 
This is largely due to a lack of understanding of the patterns and characteristics of these datasets, especially those relevant to data reduction and compressibility. 

This paper presents the first, exhaustive analysis to date of PTM datasets on storage compressibility. Our analysis spans different types of data reduction and compression techniques, from hash-based data deduplication, data similarity detection, to dictionary-coding
compression. Our analysis explores these techniques at three data granularity levels, from model layers, model chunks, to model parameters. 
We draw {new} observations that indicate that modern data reduction tools are not effective when handling PTM datasets. There is a pressing need for new compression methods that
take into account PTMs' data characteristics for effective storage reduction.  

Motivated by our findings, we design {\alg}, a simple yet effective, error-bounded, lossy floating-point compression method. {\alg} transforms floating-point parameters in such a way that the common exponent field of the transformed parameters can be completely eliminated to save storage space. 
We develop {\wf}, a compression framework that integrates {\alg} along with several other data reduction methods. {\wf} uses the most effective method to compress PTMs that exhibit different patterns. {Evaluation shows that {\wf} achieves an overall compression ratio of $1.52\times$, which is $1.31\times$, $1.32\times$ and $1.29\times$ higher than a general-purpose compressor (zstd), an error-bounded lossy compressor (SZ3), and the uniform model quantization, respectively, with negligible model accuracy loss. }

\end{abstract}

\maketitle

\settopmatter{printfolios=true}

\input{intro}
\input{related}

\input{datasets}
\input{size_analysis}

\input{compressibility_analysis}
\input{design}
\input{evaluation}
\input{conclusion}

\clearpage
\newpage

\bibliographystyle{ACM-Reference-Format}
\bibliography{refs}

\end{document}

%% file: intro.tex
\vspace{-7pt} 
\section{Introduction}
\label{sec:intro}

\begin{figure}[t]
    \centering
    \includegraphics[width=0.43\textwidth]{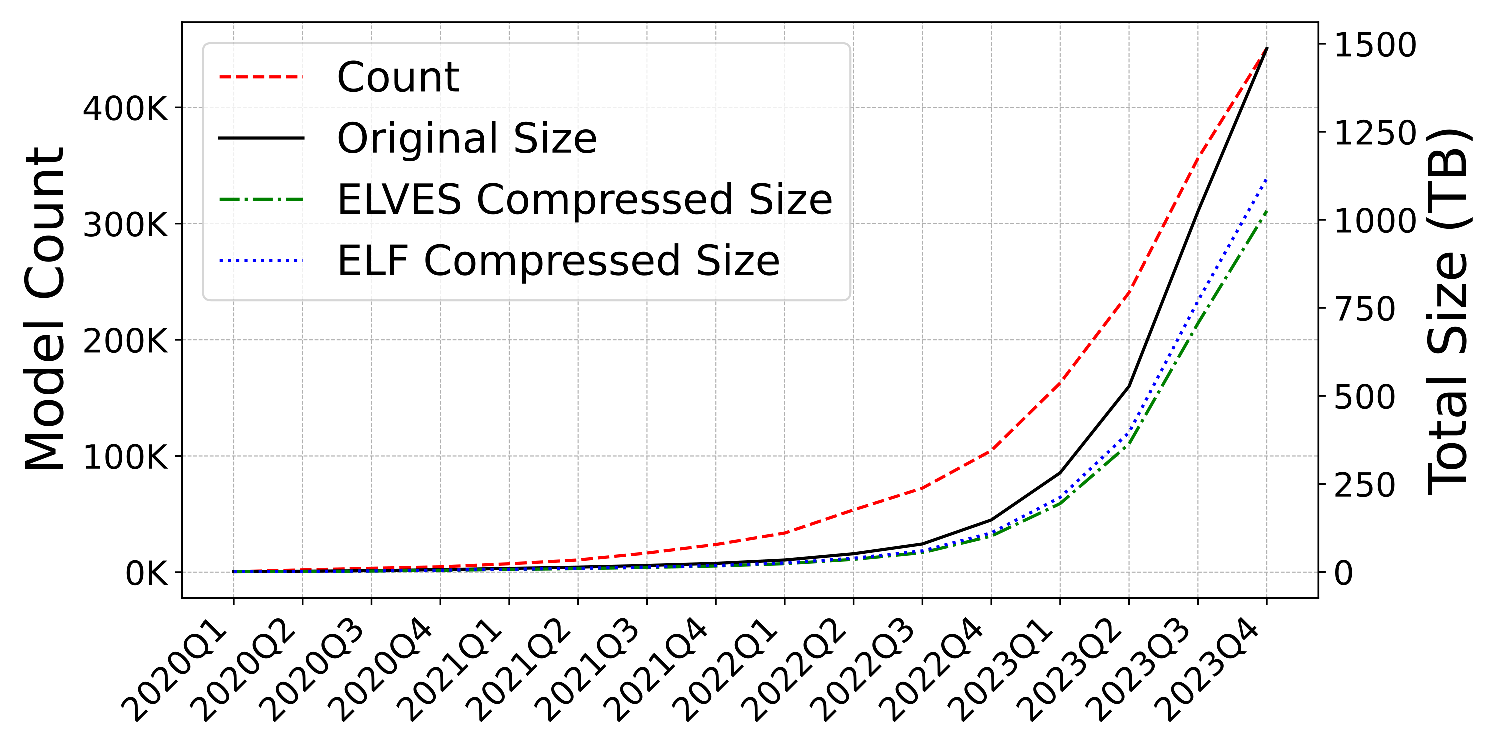}
    \vspace{-15pt}
    \caption{{Increasing trend of model count (left $Y$) and aggregate storage size (right $Y$) of Hugging Face. 
    \textit{\textmd{``{\textsf{{\wf} Compressed Size}}'' or ``{\textsf{{\alg} Compressed Size}}'' represents the storage trend after {\wf} alone or {\alg} alone is applied to Hugging Face's PTM storage.} 
    }}}
    \label{fig:trend_plot}
    \vspace{-16pt} 
\end{figure}

As artificial intelligence (AI) and machine learning (ML) continue to evolve at a fast pace, a plethora of diverse models are being created and refined. These models
reveal several emerging trends.
First, the sheer number of pre-trained models (PTMs) is skyrocketing. These models are \emph{pre-trained} to achieve a desirable accuracy for numerous tasks. PTMs are further reused to build task-specific models, which are fine-tuned with expertise~\cite{gpt_finetune_arxiv22, finetune18}. Typically, PTM datasets are 
persistently stored and managed in the format of \emph{files} by model registry services such as Hugging Face~\cite{hugging_face} and TensorFlow Hub~\cite{tensorflow_hub} to facilitate model sharing.
For example, 
Hugging Face hosts over \textcolor{black}{$450K$ PTMs (1,486.72~TB in size) as of December 31, 2023}, and this number has been growing exponentially, as shown in Figure~\ref{fig:trend_plot}. 

Second, the exponential growth of training datasets and the vast range of problem domains lead to more complex model architectures, enriched features, a significant rise in the number of parameters, and as a result, increasingly large model sizes~\cite{llama, bloom_arxiv23}. {As of the \textcolor{black}{fourth} quarter of 2023,  stored PTM datasets in Hugging Face have already exceeded \textcolor{black}{1,400~TB} (Figure~\ref{fig:trend_plot}) and the need for storage is projected to continue in the foreseeable future.}
These trends impose huge storage requirements for MLOps to store PTMs. 

The extensive storage requirements associated with large pre-trained ML models could theoretically be alleviated through data reduction techniques. These techniques include general-purpose lossless compression algorithms~\cite{gzip, zstd}, data deduplication methods for enterprise storage systems~\cite{ddfs_fast08, deduptradeoffs_fast15}, and floating-point compression algorithms for scientific datasets~\cite{zfp, sz3_icde21}, 
{time series (TS) datasets~\cite{chimp_vldb22, gorrila_vldb15, blalock2018sprintz, liu2021decomposed}, and data lakes~\cite{btrblocks_sigmod2023}}. 
However, these strategies are not effective when dealing with PTM datasets becasue:
(1)~none of existing techniques are \emph{aware of the data patterns of PTMs};
and (2)~there is \emph{a lack of understanding in those patterns,
especially those relevant to data reduction and storage compressibility}.  

{Model pruning and quantization techniques~\cite{pruning_quantization_survey_liang2021pruning, pruning_quantization_latency_hawks2021ps, pruning_quantization_accelarate_chang2021mixed}, on the other hand, are typically used for reducing the memory and computational requirements during training or inference,
posing many constraints and challenges in model integrity, usability, and accuracy when applied to PTM storage reduction.} 

To fill this gap, we present, to the best of our knowledge, the first comprehensive study of a large, real-world dataset of PTMs collected from Hugging Face on PTM storage compressibility.  
Our analysis seeks {\bf comprehensiveness} in three dimensions.
\begin{itemize}[noitemsep,leftmargin=*]
\item {\bf Scale.} We collected a total of $8,238$ PTMs from Hugging Face, which include over $75K$ files and occupy around 13~TB storage capacity. We performed the analysis on file formats, file storage footprint, and model sizes under different model categories.
\item {\bf Data reduction techniques.} We sampled a representative set of {900} models from our large-scale dataset and performed an in-depth, what-if analysis of various widely used data reduction techniques. We first studied hash-based deduplication at the model layer level and chunk level. Then, we proceeded with data similarity detection to see whether model layers or model chunks are highly similar. Finally, we examine the dictionary coding compression technique at the parameter level.  
\item {\bf Data granularity.} Our analysis explored the aforementioned techniques at three data granularity levels: model layers, model data chunks, and model parameters. 
\end{itemize}

\noindent Our multifaceted analysis induces the following {\bf observations}.
\begin{itemize}[noitemsep,leftmargin=*]
\item {\bf Real-world PTMs are large and deep.} Our analysis shows that $65.97\%$ of the $8,238$ models fall within the size range between 100~MB and 1,000~MB. {$45.83\%$ of models have 150-250 layers.} 
\item {\bf Parameters of these models are highly concentrated.} The main contents of PTMs, model parameters, are predominantly floating-point values, and most of these floating-point parameters are {\small\texttt{float32}}. Across all {900} sampled models, {$98.91\%$} of all parameters fall within the range of $(-1,1)$. 
\item {\bf Most model layers and chunks are non-duplicate, nor are they similar.} {Our analysis reveals that duplicate layers constitute a mere 5.72\% of the overall storage footprint. These layers are found across all different model categories, with a significantly higher duplication ratio in layers composed of integer-typed parameters as opposed to floating-point parameters. Our data similarity detection algorithm uncovers that only 7.98\% of all 512~B model data chunks bear any resemblance to each other.}
\item {\bf A majority of models exhibit modest-to-high parameter redundancy.} About {$48.94\%$} of models have at least $50\%$ of their parameters repeated at least once. However, widely used general-purpose dictionary coding compressors are generally not effective due to the limited length of floating-point parameters and the long distance between duplicate parameters. Nonetheless, dictionary coding is effective for {$11.56\%$} of models, where over $99\%$ of their parameters are duplicated. 
\end{itemize}

A high-level {\bf ramification} of many of these observations is \emph{a previously undisclosed insight: PTM storage compression is very challenging and existing techniques are generally ineffective due to the randomness of PTMs.} 
There is a need for new compression methods that \emph{account for PTMs' data characteristics} in order to extract most of the compressibility from the datasets. 

This paper makes the following {\bf contributions}. 
\begin{itemize}[noitemsep,leftmargin=*]
\vspace{-2pt} 
\item We conduct the first, exhaustive study of the storage compressibility of real-world pre-trained ML model datasets {and make key observations that motivate the design of new compression methods for PTM storage}. 

\item  We propose {\alg} (Exponent-Less Float-point encoding),
a new, error-bounded, lossy floating-point compression method motivated by the observations from our analysis. The idea of {\alg} is simple yet effective: since most
parameters in PTMs are within 
$(-1,1)$, {\alg} maps all parameters $\in (-1,1)$ to $[1,2)$ so that the common exponent field {\small\texttt{0b}}01111111 can be completely eliminated to save storage space. 
{{\alg} is easily parallelizable and has fast compression and decompression speed.}

\item We develop {\wf}, {an offline}  
compression framework for efficient PTM storage. 
{\wf} incorporates {\alg} along with 
hash-based deduplication, {length-distance dictionary coding}, 
and a general-purpose lossless compressor. 
Our hybrid approach collectively compresses PTM datasets that exhibit different data patterns. 

\item We develop a validation framework that generates random inputs to validate the accuracy of {\wf}-decompressed models at scale. 

\item Experimental results show that: 
{(1)~{\wf} achieves the highest compression ratio ({1.52}) for our collected dataset compared to a wide range of {11 compression methods} 
with negligible model accuracy loss;
and (2)~{in terms of compression and decompression speed, {\alg} outperforms all the 13 selected 
baseline methods}.} 
\vspace{-2pt} 
\end{itemize}

{A CR of {1.52} might not appear remarkable if viewed in isolation. However, this result is significant compared with the state-of-the-art compression methods, a factor of {$1.29\times$} improvement compared to the best baseline compressor zfp.  
A {$34\%$} reduction in storage will translate to a cost reduction of hundreds of TBs of storage hardware: if we apply {\wf} (or {\alg}) to Hugging Face's PTM storage, it could have saved \textcolor{black}{509 TB (369 TB) of storage by end of 2023 Q4} as shown in Figure~\ref{fig:trend_plot}. 
The saved storage also means a potential improvement in datacenter TCOs, encompassing benefits such as reduced cooling, lower energy usage, and decreased carbon footprint~\cite{dc_energy_link}.}

%% file: related.tex
\vspace{-4pt}
\section{Related Work}
\label{sec:related}

{\noindent\textbf{Model Pruning and Quantization.}
There is a large body of research focusing on reducing the \emph{\bf memory and computational requirement} of ML models for \emph{online} tasks such as model serving~\cite{liangpruning_nc21, deepcompression_iclr15, prune_nips15:10.5555/2969239.2969366, nodeprune_icassp14:6853595, hashnet_icml15:chen2015compressing, convlr_nips14:10.5555/2968826.2968968, gupta2022complexity, zadeh2020gobo, eisenman2022check, zhang2021qd}.}

\begin{itemize}[noitemsep,leftmargin=*]
\item{\bf {Why pruning may not be ideal for PTM storage reduction:}}
{Pruning removes insignificant layers or connections in the model, resulting in a smaller representation of the same model~\cite{prune_nips15:10.5555/2969239.2969366}. Thus, pruning impacts the integrity of the stored PTM datasets. Pruning can be generally categorized into two main types: structured pruning~\cite{structured_pruning_anwar2017structured, structured_pruning_wang2019structured} and unstructured pruning~\cite{unstructured_pruning_liao2023can, unstructured_pruning_vahidian2021personalized, gupta2022complexity}. Structured pruning may remove entire channels, filters, or layers, leading to significant model size reduction. However, structured pruning impacts the integrity of PTMs:
(1)~from a model provider perspective (e.g., Hugging Face), such irreversible changes are often not acceptable to those who share PTM datasets via model registries;
(2)~these changes to model structures may affect subsequent MLOps operations, for instance, applying models to new datasets or tasks relying on models' original structures. To end this, we include unstructured pruning, specifically global magnitude pruning~\cite{gupta2022complexity} as a baseline in Section~\ref{sec:eval}.}

\item{\bf {Why quantization may not be ideal for PTM storage reduction:}}
{Quantization~\cite{deepcompression_iclr15, krishnamoorthi2018quantizing, zadeh2020gobo, eisenman2022check, zhang2021qd} involves representing the parameters and activations of a model using fewer bits than the original data type representations, for example, converting {\small\texttt{float32}} to {\small\texttt{float16}}, {\small\texttt{int8}}. Quantization may lead to exceeding information loss, and ultimately, model accuracy loss~\cite{zeng2022glm130b}. Quantization methods can be generally categorized into quantization-aware training~\cite{quantization_training_park2018value}, dynamic quantization~\cite{quantization_dynamic_liu2022instance}, and post-training quantization~\cite{quantization_post_training_nahshan2021loss, zadeh2020gobo, zhang2021qd}. The first two are typically applied during training and inference, and are not directly applicable to PTM storage, since the storage phase does not involve any training and inference processes. For post-training quantization methods, the most popular approach currently involves converting high-precision parameters (e.g., float32) to lower-precision formats (e.g., float16, int8, or 4-bit or 3-bit representation). However, mapping virtually infinite, continuous values to a set of discrete values introduces non-negligible quantization errors. These errors accumulate through the neural network, potentially leading to a significant deviation from the original model. Furthermore, quantized models cannot use re-training or fine-tuning to ``regain'' the information loss for PTM storage, which further 
affects the models' accuracy, consequently diminishing their overall usability.}

\end{itemize}

{{\alg} is fundamentally different from pruning and quantization. 
{\alg} supports both compression and decompression, thus, is capable of preserving model structures and recovering model parameters to their original data type, though with bounded loss. In contrast, pruning and quantization are one-way processes, meaning that once a model has undergone pruning and quantization, it cannot be fully recovered to its original state due to lack of decompression. That is, quantization and pruning have irreversible effects on model information, therefore hindering any subsequent operations on PTMs, such as fine-tuning.}
Due to these drawbacks, these two techniques do not serve as ideal solutions for reducing the \emph{\bf storage requirement} of persistently-archived PTMs~\cite{hugging_face, tensorflow_hub}.

\noindent\textbf{Data Reduction and Compression.}  
Large-scale enterprise and cloud storage systems often rely on data deduplication~\cite{deduptradeoffs_fast15, dbdedup_sigmod17, ddfs_fast08} and delta compression~\cite{dedup_systor09, finesse_fast19, deltadedup_hotstorage12} to reduce storage costs, as these data---documents, source code, binary executables,  webpage objects, and more---typically show high duplication rates or are highly similar. 
General-purpose lossless compressors can reduce file sizes by identifying redundant information and representing them in a more compact form~\cite{lz_tit77, gzip, zstd, snappy}. However, these data reduction techniques are not designed to handle floating-point-based datasets, which renders them largely ineffective for PTM datasets. 
{Floating-point compression techniques for TS datasets~\cite{chimp_vldb22, gorrila_vldb15, fpc_dcc07, liu2021decomposed, blalock2018sprintz} exploit the temporal data patterns and redundancies of TS data and use delta compression or XOR operations on successive values to eliminate redundant information or resulting XOR'ed zeros for space savings.} {Columnar storage formats~\cite{btrblocks_sigmod2023, apache_parquet} are designed to compress large column datasets efficiently in data lakes by utilizing data-reduction and compression techniques, such as dictionary encoding, bit packing, and novel floating-point encoding.}
Lossy floating-point compression methods for scientific datasets (e.g., visualizations), such as SZ3~\cite{sz3_icde21} and zfp~\cite{zfp}, encode floating-point values by leveraging correlations among values. However, these floating-point compressors are not effective when it comes to PTM datasets since model parameters are cluttered, making it impossible to extract correlations or patterns.

%% file: datasets.tex
\section{Dataset Overview}
\label{sec:datasets}

\noindent\textbf{Full Dataset.}
We have downloaded pre-trained ML models from a total of $8,238$ Hugging Face repositories as of October 20, 2022. These repositories include $75,871$ files, accounting for around 13.2~TB of storage space. 
{\small\texttt{.json}} files represent the largest proportion, comprising approximately $42.5\%$ of all files, as {\small\texttt{.json}} files are predominantly used as configuration files, e.g., ``{\small\texttt{config.json}}''. {\small\texttt{.bin}} files account for $16.5\%$ of all files and primarily contain the binary data of models. Regarding the size distribution of different file formats, it is evident that the {\small\texttt{.bin}} files, which store PTMs, 
occupy the largest portion ($71.9\%$) of the storage footprint. 
We obtained category tag information from each model repository. Based on this tag information, we categorized all collected models into six categories as shown in Table~\ref{tbl:model_size_dataset1_table}. Out of all the $8,238$ models, $75.5\%$ are NLP models, consuming $78.82\%$ of the storage size, while only $2.37\%$ of models are from the CV category. {This distribution, albeit surprising, is understandable considering the increasing popularity of language models and generative AIs~\cite{cao2023comprehensive, openai_research} and their considerable size relative to other ML model types~\cite{gpt4_tr}.}

\begin{table}[t]
\caption{Distribution of model categories (full dataset). 
\textit{\textmd{NLP: natural language processing.
CV: computer vision.
RL: reinforcement learning.
Uninformed: models with no category tag information.}}
}
\vspace{-10pt}
\centering
\scalebox{0.9}{
\begin{tabular}{crr} 
\hline 
\textbf{Category} & \textbf{Count (\%)} & \textbf{Total Size in GB (\%)} \\ 
\hline 
\textbf{NLP} & 6,220 (75.5\%) & 7,661.22 (78.82\%) \\ 
\textbf{Audio} & 430 (5.22\%) & 466.79 (4.8\%) \\
\textbf{Multimodal} & 394 (4.78\%) & 358.6 (3.69\%) \\
\textbf{CV} & 195 (2.37\%) & 134.62 (1.39\%) \\
\textbf{RL} & 1 (0.01\%) & 0.0062 (0.0001\%) \\
\textbf{Uninformed} & 998 (12.12\%) & 1,098.5 (11.3\%) \\ 
\textbf{Overall} & 8,238 (100\%) & 9,719.73 (100\%) \\
\hline
\end{tabular}
}
\label{tbl:model_size_dataset1_table}
\vspace{-10pt}
\end{table}

\noindent\textbf{Sampled Dataset.}
We observe from the full dataset that {\small\texttt{.bin}} files that store the binary data of PTMs predominantly occupy the storage footprint, therefore, we focus on examining the characteristics and compressibility of these binary data throughout the rest of the paper. 
To do so, we use smaller samples of the full dataset that can fit within the storage capacity of typical storage server machines. 
{The sampled dataset features a more balanced distribution of model categories to avoid bias (Table~\ref{tbl:model_size_dataset2_table}). This dataset includes 150 models each from the Audio, Multimodal, CV, and Uniformed categories. 
The NLP category contains 300 models, reflecting its prominence and prevalence in current applications. 
Unless stated otherwise, the rest of the paper will be focused on the 900-model, sampled dataset.}  

\begin{table}[t]
\caption{{Distribution of model categories (sampled dataset).}
}
\vspace{-10pt}
\centering
\scalebox{0.9}{
\begin{tabular}{crr} 
\hline
\textbf{Category} & \textbf{Count (\%)} & \textbf{Total Size in GB (\%)} \\ 
\hline 
\textbf{NLP} & {300 (33.33\%)} & {170.85 (29.67\%)} \\  
\textbf{Audio} & {150 (16.67\%)} & {154.30 (26.79\%)}  \\
\textbf{Multimodal} & {150 (16.67\%)} & {97.81 (16.99\%)} \\
\textbf{CV} & {150 (16.67\%)} & {58.74 (10.20\%)} \\
\textbf{Uninformed} & {150 (16.67\%)} & {94.18 (16.35\%)} \\
\textbf{Overall} & {900 (100\%)} & {575.88 (100\%)} \\
\hline
\end{tabular}
} 
\label{tbl:model_size_dataset2_table}
\vspace{-10pt} 
\end{table}

%% file: size_analysis.tex
\section{Analysis: Sizes and Contents}
\label{sec:size_content_analysis}

\begin{figure*}[t]
\vspace{-12pt}
\begin{center}
\begin{minipage}{\textwidth}
\begin{minipage}[b]{0.322\textwidth}
\includegraphics[width=1\textwidth]{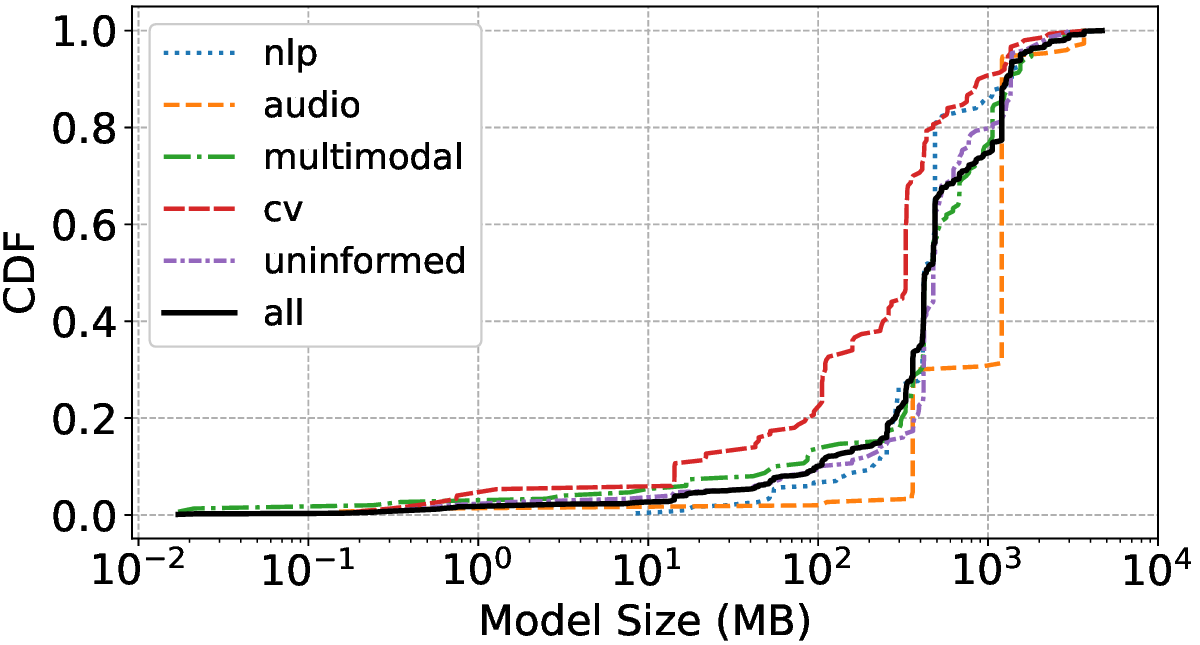}
\vspace{-20pt}
\caption{{Model size distribution of the sampled dataset.}}
\label{fig:model_size_log_cdf_900}
\end{minipage}
\hspace{.5pt}
\begin{minipage}[b]{0.322\textwidth}
\includegraphics[width=1\textwidth]{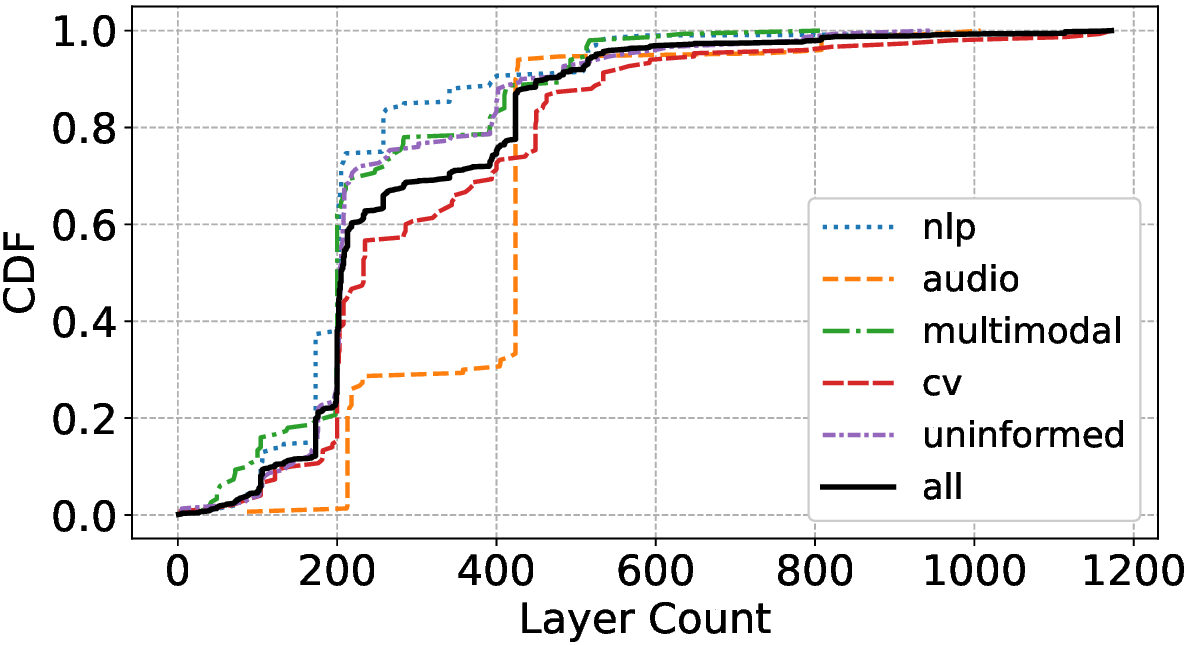}
\vspace{-20pt}
\caption{{Model layer number CDF for different categories.}}
\label{fig:layer_number_cdf}
\end{minipage}
\hspace{.5pt}
\begin{minipage}[b]{0.322\textwidth}
\includegraphics[width=1\textwidth]{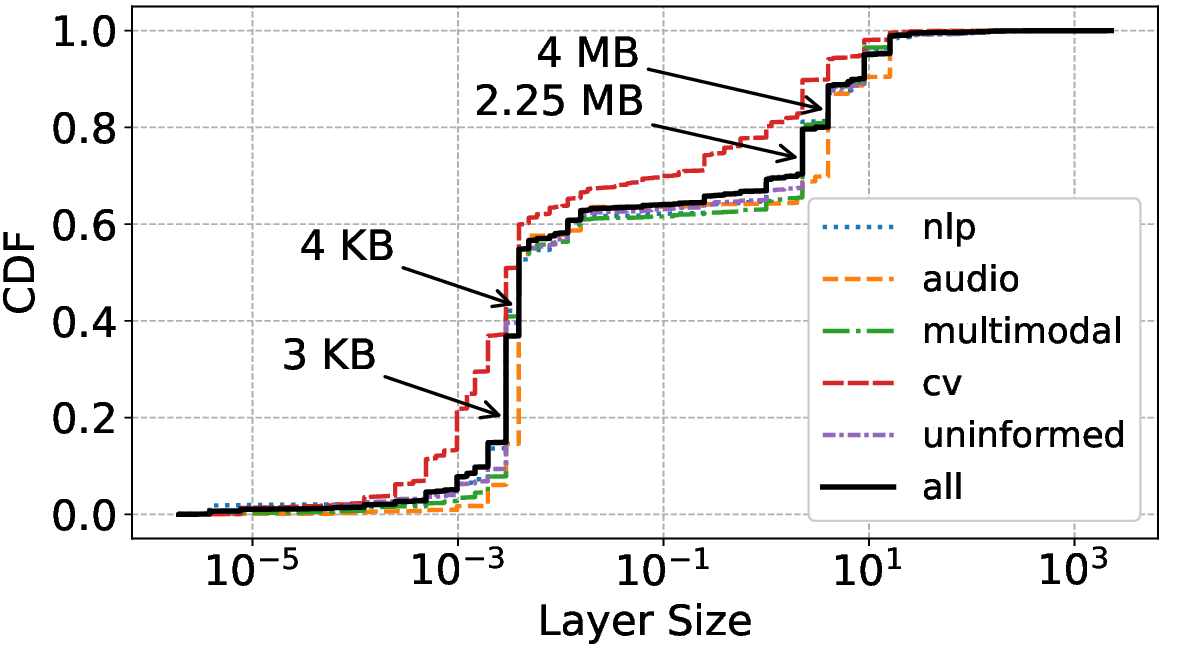}
\vspace{-20pt}
\caption{{Model layer size CDF for different categories.}}
\label{fig:layer_size_log_cdf}
\end{minipage}
\end{minipage}
\end{center}
\vspace{-10pt}
\end{figure*}

This section presents our model size and content analysis that aims to answer the following research questions (RQs):
\begin{itemize}[noitemsep,leftmargin=*]
\item{\bf RQ1:} What are the sizes of PTMs in different categories?
\item{\bf RQ2:} What are the layer counts and 
sizes in these models?
\item{\bf RQ3:} What types of data (parameters) are stored in these models?
\end{itemize}

\noindent\textbf{Model Sizes.}
{We first analyze model sizes. 
Figure~\ref{fig:model_size_log_cdf_900} shows that the size distributions for different model categories exhibit \emph{similar} trends. Specifically, 64.78\% of models fall within the size range between 100~MB and 1,024~MB, with an additional 25.22\% surpassing the 1~GB threshold. Counting all models, the \{$50\%$-ile, $75\%$-ile, $90\%$-ile\} are \{433~MB, 1,036~MB, 1,281~MB\}, highlighting the substantial file size footprint characteristic of PTMs. 
}

\noindent\textbf{Model Layers.} 
We extracted the layer information of all the {900} models in the sampled dataset and reported the layer count and layer size statistics in Figure~\ref{fig:layer_number_cdf}-\ref{fig:layer_size_log_cdf}. 
We make the following observations:
(1)~{Around {$75\%$} of models have over 200 layers}, likely because most models need a deep structure to capture complex feature information (Figure~\ref{fig:layer_number_cdf}). 
(2)~Across all categories, 
a majority of audio models are deep, with {$70\%$} having more than 400 layers (Figure~\ref{fig:layer_number_cdf}). 
(3)~As shown in Figure~\ref{fig:layer_size_log_cdf}, layer sizes exhibit a step-like distribution, with {$57.84\%$} of layer sizes concentrated around the scale of 3~KB, 4~KB, 2.25~MB, and 4~MB. 

\begin{table}[htbp]
\centering
\caption{{Model layer data type distribution.}
\textit{{\textmd{Others include  {\small\texttt{float64}}, {\small\texttt{uint8}}, and {\small\texttt{int64}}.}}}}
\vspace{-10pt}
\scalebox{0.75}{
\begin{tabular}{crrrr} 
\hline 
\textbf{Layer Type} & \textbf{Count (\%)} & \textbf{Total Sz in GB (\%)} & \textbf{Avg Para \#} & \textbf{Avg Sz in MB} \\ 
\hline 
\textbf{float32} & {240,966 (96.95\%)} & {557.84 (96.87\%)} & {621,421} & {2.37} \\ 
\textbf{float16} & {4,018 (1.62\%)} & {14.51 (2.52\%)} & {1,939,421} & {3.70} \\
\textbf{others} & {3,561 (1.43\%)} & {3.53 (0.61\%)} & {595,181} & {1.02} \\
\textbf{Overall} & {248,545 (100\%)} & {575.88 (100\%)} & {642,352} & {2.37} \\
\hline
\end{tabular}
}
\label{tbl:layer_type_table}
\vspace{-15pt}
\end{table}

\noindent\textbf{Model Data Types.}
Next, we examine the data types of the sampled model dataset. We find that all parameters within an individual layer are of the same data type. Thus, we break down all model layers by data types and report the statistics of layer data types in Table~\ref{tbl:layer_type_table}. 
{Most layers are of {\small\texttt{float32}} type with around $97\%$ in both count and storage footprint.} 
{Layers of type {\small\texttt{float16}} and others make up 2.52\%, and 0.61\% of the total storage space, respectively. }

\begin{figure}[t]
    \centering
    \includegraphics[width=0.38\textwidth]{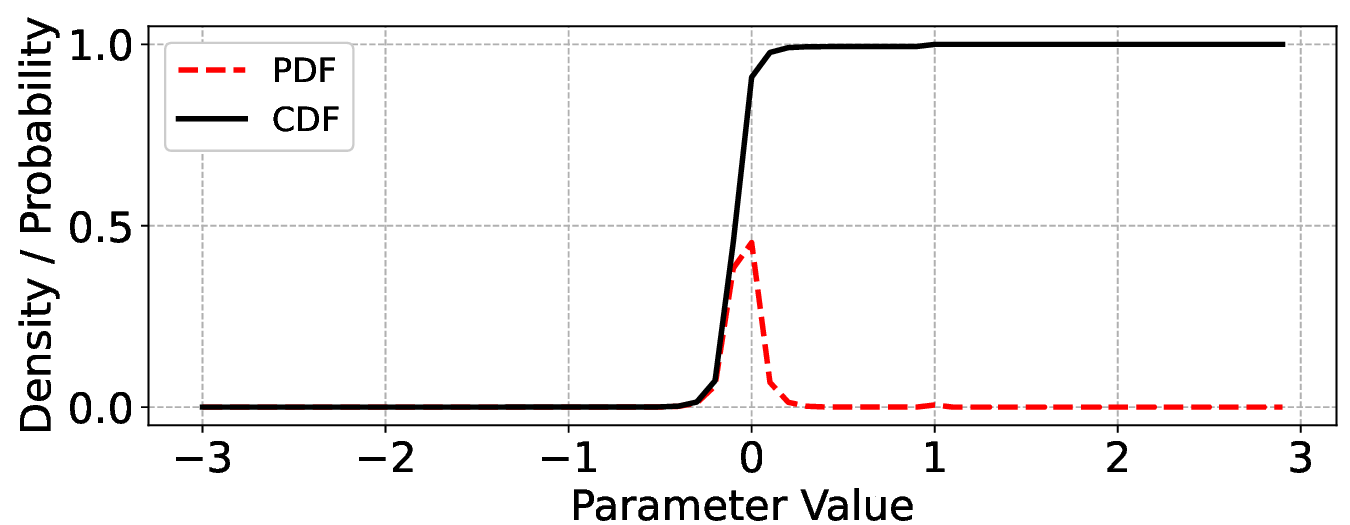}
    \vspace{-10pt}
    \caption{{Parameter value distribution.}}
    \label{fig:parameter_value_cdf_pdf}
    \vspace{-15pt}
\end{figure}

\noindent\textbf{Model Parameters.}
We then analyze the distribution of parameter values.
As shown in Figure~\ref{fig:parameter_value_cdf_pdf},  {$98.91\%$} of all parameters fall within the range of $(-1, 1)$, with {$50.50\%$} in the interval $(-1, 0]$ and {$48.41\%$} in the interval $(0, 1)$.

\vspace{-5pt}
\begin{tcolorbox}[breakable,width=0.48\textwidth,title={Implications},boxrule=.3mm,colback=white,coltitle=white,left=.5mm, right=.5mm, top=.5mm, bottom=.5mm]    
   \emph{$\bullet$ Real-world, pre-trained ML models are considerably large in size. The rapidly increasing number of ML models poses significant challenges to MLOps for storing and managing the exponentially increasing volume of model datasets. \\
    $\bullet$ These models typically contain a massive number of layers, potentially providing intriguing opportunities for data reduction using layer-based (or chunk-based) deduplication techniques and/or similarity-based delta compression techniques.\\ 
    $\bullet$ Parameter values are concentrated within a narrow range of $(-1, 1)$, implying potential opportunities for the application of 
    compression/encoding methods~\cite{lz_tit77, gzip, gorrila_vldb15, zfp, sz3_icde21, fpc_dcc07}. These compression methods use a combination of techniques including prediction~\cite{sz3_icde21, lossy_bigdata18}, XOR-based zero encoding~\cite{gorrila_vldb15, fpc_dcc07}, and error-encoding~\cite{lossy_hpdc20} to compress floating-point model datasets.  
}
\end{tcolorbox}

%% file: compressibility_analysis.tex
\section{Analysis: Compressibility}
\label{sec:compressibility_analysis}

In this section, we present our two-dimensional, what-if analysis of model data reduction and compressibility. Along the \emph{data granularity dimension}, our analysis explores the potential data reduction yield at three different levels: model layers, data chunks, and individual parameters. Along the \emph{data reduction and compression technique dimension}, we consider three widely used techniques: hash-based deduplication~\cite{deduptradeoffs_fast15}, similarity-based delta compression~\cite{finesse_fast19}, and distance-encoding~\cite{lz_tit77}. 
\emph{Note that data reduction techniques integrated within storage systems are closely tied to the specific systems. It is difficult to directly extract these techniques and compare them in our analysis. As such, we implemented several representative techniques for analysis and comparison purposes.}
Specifically, our analysis aims to answer the following RQs:
\begin{itemize}[noitemsep,leftmargin=*]
\item{\bf RQ1:} Does duplication exist among model layers or model data chunks? If so, would data deduplication help in reducing the model data storage footprint? 
\item{\bf RQ2:} Are there any model layers or chunks that are highly similar?  
\item{\bf RQ3:} What is the repetition pattern of model parameters? If parameter-level repetition exists, how can this redundancy be eliminated or mitigated? 
\end{itemize}

\begin{table}[htbp]
\centering
\vspace{-10pt}
\caption{{Model layer duplication statistics based on data types.}
\textit{{\textmd{The percentages in Columns 3 and 5 represent the proportion of the total count and total size under that row.}}}
}
\vspace{-10pt}
\scalebox{0.88}{
\begin{tabular}{crrrr} 
\hline 
\textbf{Layer Type} & \textbf{Count} & \textbf{Dup \%} & \textbf{Total Sz in GB} & \textbf{Dup Sz in GB (\%)} \\ 
\hline 
\textbf{float32} & {240,966} & {8.35\%} & {557.84} & {30.14 (5.40\%)} \\
\textbf{float16} & {4,018}   & {3.61\%} & {14.51} &  {0.14 (0.96\%)} \\
\textbf{float64} & {199}     & {0\%}    & {0.81} &  {0 (0\%)} \\
\textbf{uint8}   & {1,597}   & {99.81\%} & {1.75} &  {1.74 (99.43\%)} \\
\textbf{int64}   & {1,765}   & {96.77\%} & {0.97} &  {0.94 (96.91\%)} \\
\textbf{Overall} & {248,545} & {9.48\%} & {575.88} &  {32.96 (5.72\%)} \\
\hline
\end{tabular}
}
\label{tbl:dup_outside_table}
\vspace{-10pt} 
\end{table}

\vspace{-4pt} 
\subsection{Model Layer- and Chunk-level Duplication}
\label{subsec:dedup}

\noindent\textbf{Does Duplication Exist among Model Layers?} 
Hash-based data deduplication method partitions the target dataset into fine-grained chunks, computes the hash values (i.e., fingerprints) of all data chunks, scans the partitioned dataset, and performs deduplication by removing duplicate chunks with identical fingerprints to save storage space.  
To understand if there is any duplication among model layers, we scanned the {900} models in the sampled dataset to compute the layer fingerprints. 
Our results are discouraging. {The analysis reveals that a mere 5.72\% of the total storage footprint for model layers is accounted for by duplication (see Table~\ref{tbl:dup_outside_table})}. For {\small\texttt{float32}}-typed layers, duplicate layers only occupy {30.14~GB} out of a total of {557.84~GB}. Although layers of type {\small\texttt{uint8}} and {\small\texttt{int64}} are largely duplicate, their overall fraction is negligible.

\begin{table}[htbp]
\centering
\caption{{Distribution of duplicate chunks based on data type.
\textit{\textmd{Columns 3 and 4 display the size of duplicate chunks using fixed-size chunking (FSC) of 4~KB and 512~B, and Column 5 indicates the total size of duplicate chunks determined by content-defined chunking (CDC) with chunk sizes ranging from 128~B to 128~KB.}}}}
\vspace{-10pt}
\scalebox{0.85}{ 
\begin{tabular}{crrrr} 
\hline 
\multirow{2}{*}{\textbf{Data Type}} &  \multirow{2}{*}{\textbf{Total Sz (GB)}} & \multicolumn{3}{c}{\textbf{Size of Duplicates in GB (\%)}} \\
\cline{3-5} & & \multicolumn{1}{c}{4 KB {(FSC)}} & \multicolumn{1}{c}{512 B {(FSC)}} & \multicolumn{1}{c}{{CDC}} \\
\hline   
\textbf{float32} &  {557.84}   &  {40.35 (7.23\%)}     & {42.92 (7.69\%)}  & {44.50 (8.16\%)}  \\ 
\textbf{float16} &  {14.51}    &  {0.14 (0.96\%)}     & {0.14 (0.96\%)}   & {0.15 (1.03\%)}   \\
\textbf{float64} &  {0.81}     &  {0 (0\%)}           & {0 (0\%)}         & {0 (0\%)}         \\
\textbf{uint8}   &  {1.75}     &  {1.74 (99.43\%)}    & {1.74 (99.43\%)}  & {1.74 (99.43\%)}  \\ 
\textbf{int64}   &  {0.97}     &  {0.94 (96.91\%)}    & {0.96 (98.97\%)}  & {0.96 (98.97\%)}  \\
\textbf{Overall} &  {575.88}   &  {43.17 (7.50\%)}     & {45.76 (7.95\%)}   & {47.35 (8.22\%)}   \\
\hline
\end{tabular}
}
\label{tbl:dup_chunks}
\vspace{-10pt} 
\end{table}

\noindent\textbf{Does Duplication Exist among Model Chunks?} 
{Next, we conduct chunk-based duplication analysis based on fixed-size chunking (FSC) and content-defined chunking (CDC)~\cite{cdc_quinlan2002venti} approaches at chunk granularity. Table~\ref{tbl:dup_chunks} shows the results. 
With FSC, chunks of 512~B exhibit a higher duplication ratio across the majority of data types compared to chunks of 4~KB. We utilized FastCDC~\cite{fast_cdc_xia2016fastcdc}, a widely used, state-of-the-art, gear-based CDC method, on our dataset. We find that the size of the duplicates detected by FastCDC is 47.35 GB, accounting for 8.22\% of the total dataset size. This is 9.68\% and 3.47\% higher than FSC with a chunk size of 4~KB and 512~B, respectively. 
\emph{Nevertheless, both the FSC and CDC duplication
analysis shows similarly negative results, indicating that hash-based data deduplication might not effectively reduce the storage size of PTM datasets.}}

\subsection{Model Layer- and Chunk-level Similarity}
\label{subsec:delta}

We next study whether PTMs contain data that is similar but not exactly identical. In storage systems, delta compression often complements deduplication as a data reduction technique in order to eliminate redundancy among non-duplicate yet highly similar chunks~\cite{dedup_systor09, deltadedup_hotstorage12,dbdedup_sigmod17}. 
For example, if chunk $D_2$ is similar to a base chunk $D_1$, a delta compressor will only store the differences, i.e., the \emph{delta}, and the mapping between $D_2$ and $D_1$, by removing the redundant data for improved storage efficiency. 

\noindent\textbf{How to Detect Data Chunk Similarity?}
Widely used data similarity (resemblance) detection methods compute ``super features'' (SFs)~\cite{resemblance_sequences97, rebl_atc04} based on the Rabin fingerprints~\cite{rabin_fingerprints}\footnote{Rabin fingerprints compute a group of hash values using a sliding window that slides from the start to the end of the data chunk. The size of the sliding window is configurable and is set to 48 bytes as per \cite{finesse_fast19}.} of data chunks and use the computed SFs to detect similar chunks. For example, Finesse~\cite{finesse_fast19}, a state-of-the-art method, works as follows.
(1)~The base and target data chunks, $D_1$ and $D_2$, are partitioned into four sub-chunks each, and a group of hash values based on Rabin fingerprints are computed for all eight sub-chunks.
(2)~For both $D_1$ and $D_2$, three SFs are constructed.
The first SF is constituted by using the largest hash values from each of its four sub-chunks, the second SF from the second largest hash values from the four sub-chunks, and the third SF from the third largest hash values from the four sub-chunks.
(3)~A hash value based on SFs for $D_1$ and $D_2$ is computed. If the hash value of $D_1$'s SFs is the same as that of $D_2$, it indicates that $D_1$ and $D_2$ are highly similar.

\begin{table}[htbp]
\centering
\caption{{Similarity ratios for different granularities.
\textit{\textmd{Similarity ratio is defined as the size of similar layers/chunks divided by the total size. (Note that the true similarity ratios are lower than reported in this table as the entire sampled dataset includes duplicate data.)
}}}}
\vspace{-10pt}
\scalebox{0.85}{
\begin{tabular}{crrrr} 
\hline 
\multirow{2}{*}{\textbf{Data Type}} &  \multirow{2}{*}{\textbf{Total Sz (GB)}} & \multicolumn{3}{c}{\textbf{Size of Similar Data in GB (\%)}} \\
\cline{3-5} & & \multicolumn{1}{c}{Layer} & \multicolumn{1}{c}{4 KB} & \multicolumn{1}{c}{512 B} \\
\hline  
\textbf{float32} &  {557.84}   & {30.17 (5.41\%)}     & {40.39 (7.24\%)}       & {43.12 (7.73\%)} \\ 
\textbf{float16} &  {14.51}    & {0.14 (0.96\%)}      & {0.14 (0.96\%)}        & {0.14 (0.96\%)} \\
\textbf{float64} &  {0.81}     & {0 (0\%)}            & {0 (0\%)}              & {0 (0\%)} \\
\textbf{uint8}   &  {1.75}     & {1.74 (99.43\%)}     & {1.74 (99.43\%)}       & {1.74 (99.43\%)} \\ 
\textbf{int64}   &  {0.97}     & {0.94 (96.91\%)}     & {0.95 (97.94\%)}       & {0.96 (98.97\%)} \\
\textbf{Overall} &  {575.88}   & {32.99 (5.73\%)}     & {43.22 (7.51\%)}       & {45.96 (7.98\%)} \\
\hline
\end{tabular}
}
\label{tbl:similarity}
\vspace{-10pt} 
\end{table}

\noindent\textbf{Are Model Layers or Chunks Highly Similar?}
Methods like Finesse introduce significant computational overhead. To explore the potential of delta compression, we designed and implemented a simple and efficient data similarity detection algorithm that approximates Finesse. Our approximation algorithm samples a parameter for every $N$ parameters from each model layer or model chunk, with $N$ setting to 32 in this method, computes the hashes of sampled parameters, and compares the hashes of two layers/chunks. Compared to Finesse which uses a sliding window to compute SF hashes, our algorithm reduces computational requirements by using sampling and a sliding window of 1. A side effect of the approximation is a potentially higher false positive rate, where two layers/chunks with sparse similarities might be inaccurately detected as highly similar. However, as shown in Table~\ref{tbl:similarity}, the similarity ratios for model layers and 4~KB/512~B chunks remain remarkably low. For example, for the entire dataset, only {$7.98\%$} of 512~B blocks are identified as similar. \emph{These negative results suggest that similarity-based delta compression will not be effective for reducing the storage footprint of PTMs.}

\subsection{Model Parameter-level Duplication}
\label{subsec:parameter}

\begin{figure}[t]
\begin{center}
\begin{minipage}{\textwidth}

\begin{minipage}[b]{0.23\textwidth} 
\includegraphics[width=1\textwidth]{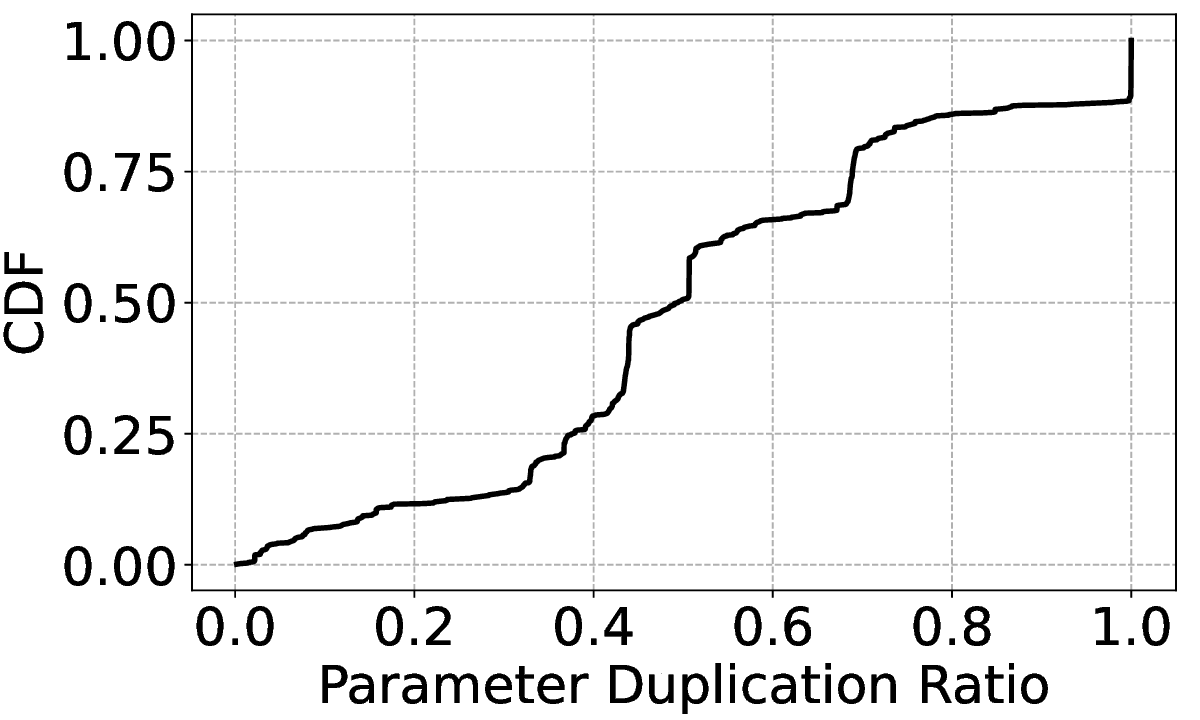}
\vspace{-15pt}
\caption{{The parameter duplication ratio of all 900 models. Each data point in the CDF curve represents a model's duplication ratio.}
}
\label{fig:para_dup_ratio}
\end{minipage}
\hspace{.5pt}
\begin{minipage}[b]{0.23\textwidth}
\includegraphics[width=1\textwidth]{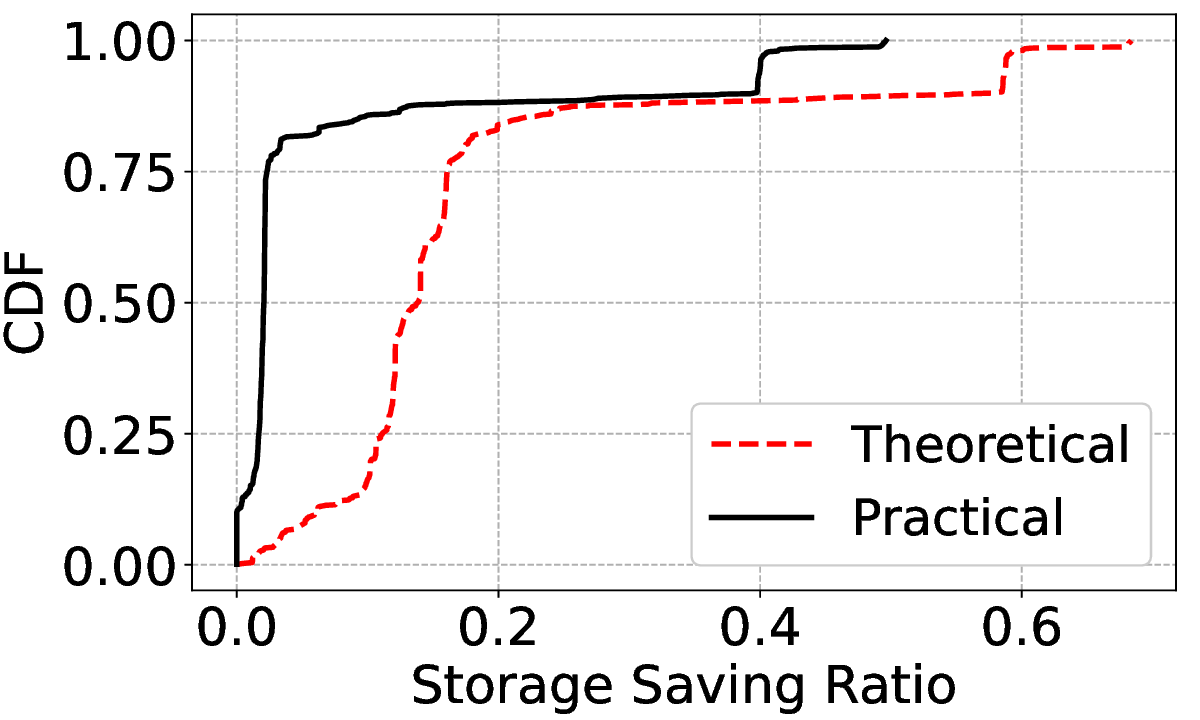}
\vspace{-15pt}
\caption{{Storage saving ratios ($\%$ storage space saved) by practical DE (with overhead included) and theoretical DE (without overhead).}
}
\label{fig:storage_saving_ratio}
\end{minipage}

\end{minipage}
\end{center}
\vspace{-15pt}
\end{figure}

\noindent\textbf{What is the Repetition Pattern of Model Parameters?}
Recall we have shown in \cref{sec:size_content_analysis} that the values of model parameters are concentrated within a small range of $(-1, 1)$. Going one step deeper, 
we study the repetition pattern of individual parameters by counting the duplication ratio of parameters for each individual model from our sampled dataset. Here the parameter duplication ratio for a model is defined as the fraction of repetitive parameters. Figure~\ref{fig:para_dup_ratio} shows that {$48.94\%$} of models exhibit a parameter duplication ratio of {over} $50\%$, meaning that these models have at least half of their parameters duplicated at least once. Interestingly, {$11.56\%$} of models have over $99\%$ duplicated parameters. The high duplication ratios imply potential opportunities for utilizing general-purpose compression
methods to reduce parameter redundancies.

\noindent\textbf{Will Off-the-Shelf Dictionary Coding Work?}
The high parameter duplication ratio motivates us to conduct a what-if analysis to study the feasibility of existing compression methods in reducing the parameter-level redundancy. 
We first explore a compression technique that is commonly used in today's general-purpose compressors, called dictionary coding~\cite{text_compression_acmj82}. Dictionary coding works by using an external macro scheme, i.e., a separately stored dictionary data structure, to maintain a mapping between duplicated sequence patterns, e.g., text strings, and codes that 
locate the sequences. Given the high duplication ratio of PTM datasets, it seems that {this might work.} However, 
we found that, while most models have duplicate parameters, the absolute amount of unique duplicate parameters, i.e., the working set size of duplicate parameters, in a model can be enormous, which makes the space complexity of the dictionary extremely large. Worse, the average repetition frequency for all duplicate parameters is on the lower end. For example, in our dataset, {$34.35\%$} of models have over $60\%$ of parameter duplicates; however, these duplicate parameters have an average repetition frequency of around {12}, with a total unique parameter count exceeding {6.8}~billion. This implies that at least 33 bits in the length code would be required to encode the whole dictionary. Given this, dictionary coding offers no data reduction benefit for PTM datasets. 

We then consider a more efficient dictionary coding variant.
{Without external macros,}
this method encodes duplicate sequence patterns in pointers. A pointer is a length-distance $(D, L)$ pair~\cite{text_compression_acmj82, lz_tit77}, where the ``distance'' $D$ tells the compressor (and the decompressor) how far back to look for the start of the repeated sequence, and the ``length'' $L$ tells the compressor how many characters make up the repeated sequence. 
Real-world implementations such as LZ77~\cite{lz_tit77} typically use a sliding window to provide a dynamic dictionary of duplicate sequences that can be referred back to.

This general-purpose compression method can be more space-efficient to store a short pointer that refers to an earlier occurrence of a string than to store the whole string again, especially if the string itself is long and/or repeated frequently. 
This method typically works well for text-based datasets~\cite{gzip, zip}. 
Dealing with floating-point datasets such as PTM datasets, however, becomes challenging due to the following reasons.
(1)~Unlike string-based text datasets where duplicate sequences can be long, a duplicate model parameter is short, e.g., the most common data type in PTM datasets---{\small\texttt{float32}}-typed parameters (Table~\ref{tbl:layer_type_table})---are only 4 bytes long. 
While the good news, in our case, is that
the length value can be omitted if we target {\small\texttt{float32}}-typed parameters only, simply because {\small\texttt{float32}} parameters are of fixed length; the bad news is that the limited length of duplicate parameters puts a hard constraint on the potential gain in storage reduction.  
(2)~Duplicate parameters are sparsely distributed within a model, requiring a large sliding window size and relatively long distance values. 
In a common implementation, the distance values might be represented as 16-bit integers, which can encode a distance of at most $2^{16} = 65,536$. Unfortunately, in most models, the distance between two adjacent duplicate parameters is longer than that we observed from our dataset. Of course, this problem can be addressed by using longer distance values. However, the longer the distance value, the less data reduction gain the compressor would achieve. In theory, we need to control the distance value length to be shorter than 4 bytes ($2^{32}$) in order to receive gains.

\begin{figure}[t]
\begin{center}
\includegraphics[width=0.45\textwidth]{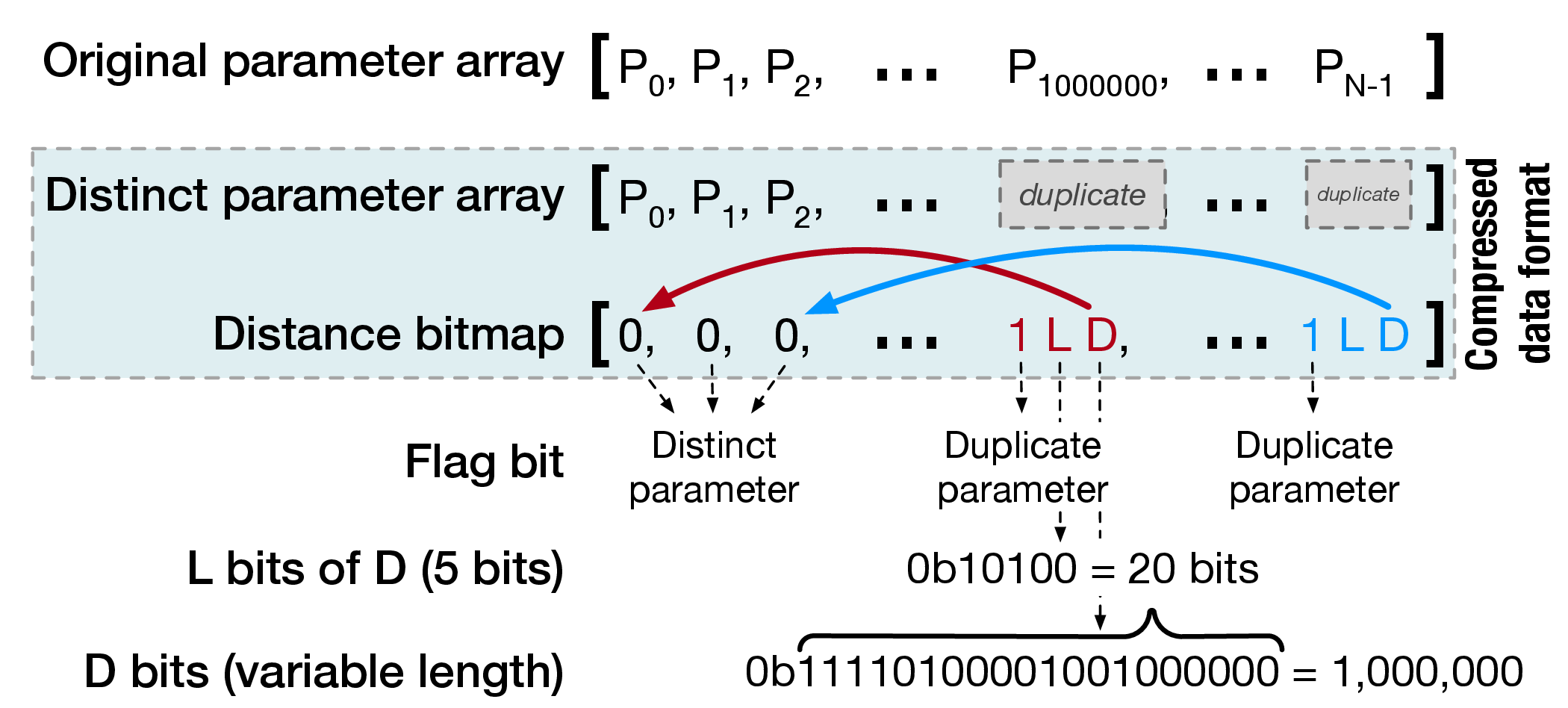}
\vspace{-12pt}
\caption{An example of distance-encoding compression. 
\textit{\textmd{L: length. D: distance.
The distinct parameter array does not store duplicate parameters 
marked in dashed-line, shady boxes. In this example, $P_{1000000}$ duplicates with $P_0$. Therefore, DE toggles $P_{1000000}$'s flag bit as 1 and adds a length-distance pair $(L,D)$. The 5-bit $L$ field encodes 20, indicating that the bit length for $D$ is 20 bits. 
$D$ encodes a decimal value of $1,000,000$, meaning that this duplicate refers back to a parameter that {is $1,000,000$ {\small\texttt{float32}} parameters ahead}.}}
}
\label{fig:de_example}
\end{center}
\vspace{-22pt}
\end{figure}

\noindent\textbf{How to Minimize Model Parameter Redundancy?}
To verify whether length-distance dictionary coding is effective, we designed and implemented an efficient length-distance-based compression method and data format, which we call distance-encoding (DE),
targeting {\small\texttt{float32}}-typed {and} {\small\texttt{float64}}-typed model parameters. 
Our DE method stores a compressed model parameter file into two logically decoupled arrays: a distinct parameter array that is used to store duplicated and unique, and non-duplicate 
parameters and a distance bitmap that is used to store metadata to keep track of the pointers for duplicate parameters. Figure~\ref{fig:de_example} gives an example of the data format. 
To compress, our DE \emph{compressor} takes a linear pass of the model dataset. DE stores distinct parameters as is in the distinct parameter array and records a flag bit of 0 in the distance bitmap. 
DE skips a duplicate parameter in the distinct parameter array and records a flag bit of 1 followed by an $(L,D)$ pair in the distance bitmap. $L$ is a fixed-length, 5-bit field, which records the bit length of the next field, distance $D$, so that \emph{the decompressor} knows how many bits to read in order to decode $D$. The 5 bits in $L$ can encode a distance of at most $2^{31}$ if all five bits are used  ($2^5-1 = 31$). As such, the $D$ field can have variable length, ranging from 1 to 31 bits. Note that the fields of flag bit and $L$ bits introduce storage overhead.

We evaluated DE on our {900}-model dataset.  Figure~\ref{fig:storage_saving_ratio} plots the storage saving ratios achieved by: (1)~a theoretical compressor without adding the flag bit and $L$-field overhead, representing a best-case baseline, and (2)~a practical compressor that includes the extra overhead. DE, in theory, can achieve at least $10\%$ storage savings for {$83.67\%$} of {900} models. However, this comes with the catastrophic consequence of being not decompressible due to the absence of metadata.
Taking into account the extra metadata overhead, DE's efficacy decreases dramatically---{it can only provide the same space savings for $13.22\%$ of the models}, which corresponds to the vertical curve at the top-right corner of Figure~\ref{fig:para_dup_ratio}. {But the good news is that the average storage saving ratio for these models is remarkable, at about {$33\%$}.}
This is because these models have over $99\%$ of parameters duplicated, representing a best-case scenario for DE to be effective.  
\emph{This mixed result suggests that 
general-purpose length-distance-based compression methods can achieve a reasonably high compression ratio only if model parameters are highly duplicated.}

\vspace{-2pt}
\begin{tcolorbox}[breakable,width=0.48\textwidth,title={Implications},boxrule=.3mm,colback=white,coltitle=white,left=.5mm, right=.5mm, top=.5mm, bottom=.5mm]    
   \emph{$\bullet$ The effects of hash-based data deduplication approaches are double-edged. First, duplicate layers in our sampled dataset make up only {$5.72\%$} of the storage size. 
   {Both the FSC-based and CDC-based deduplication see very limited duplication ratio.}
   We thus expect hash-based data deduplication approaches to be generally ineffective in reducing storage costs. Second, a much higher level of layer duplication exists in the integer-typed model layers, suggesting a potential avenue for future research. Overall, the effectiveness of data deduplication for PTM datasets appears to be minimal.\\
   $\bullet$ Our approximation data similarity detection algorithm reveals that only {$7.98\%$} of all 512~B model chunks, including duplicate chunks, bear resemblance to each other. This result {suggests} that 
   there is {limited} similarity within PTM datasets that could be leveraged by delta compression techniques.\\
   $\bullet$ Model parameter-level duplication is virtually universal---{with all 900 models in our dataset having duplicate parameters}. However, this does not imply that widely used, general-purpose compression algorithms, such as length-distance-based dictionary coding, will be effective for PTMs. Two main reasons contribute to such lack of compressibility. First, the unit of the duplicate sequence---floating point numbers---is short and most duplicate parameters repeat infrequently. Second, the distance between duplicate parameters is typically long, requiring lengthy bits to encode the distance. On a positive note, though, {$11.56\%$} of models have over $99\%$ duplicated parameters, and therefore, could benefit from high storage savings by using general-purpose compressors. 
}
\end{tcolorbox}  
\vspace{-2pt}

%% file: design.tex
\vspace{-4pt}
\section{{\alg} and {\wf} Design}
\label{sec:design}

In this section, we first introduce a new, error-bound, lossy floating-point compression method {\alg} (Exponent-Less Float-point encoding), motivated by the observations from \cref{sec:size_content_analysis} and \cref{sec:compressibility_analysis}. 
{We then present a compression framework named {\wf}, which integrates two main compression methods, {\alg} (\cref{subsec:elf}) and DE (\cref{subsec:parameter}), along with several other data reduction methods, to compress pre-trained ML model datasets. {\alg} compresses models that primarily consist of floating-point parameters that fall within the range of $(-1,1)$, while DE targets models that have a significant proportion of out-of-range parameters. These two methods complement each other, thereby maximizing overall storage efficiency.} 

\vspace{-4pt}
\subsection{The {\alg} Compression Algorithm}
\label{subsec:elf}

\begin{figure}[t]
\begin{center}
\includegraphics[width=0.46\textwidth]{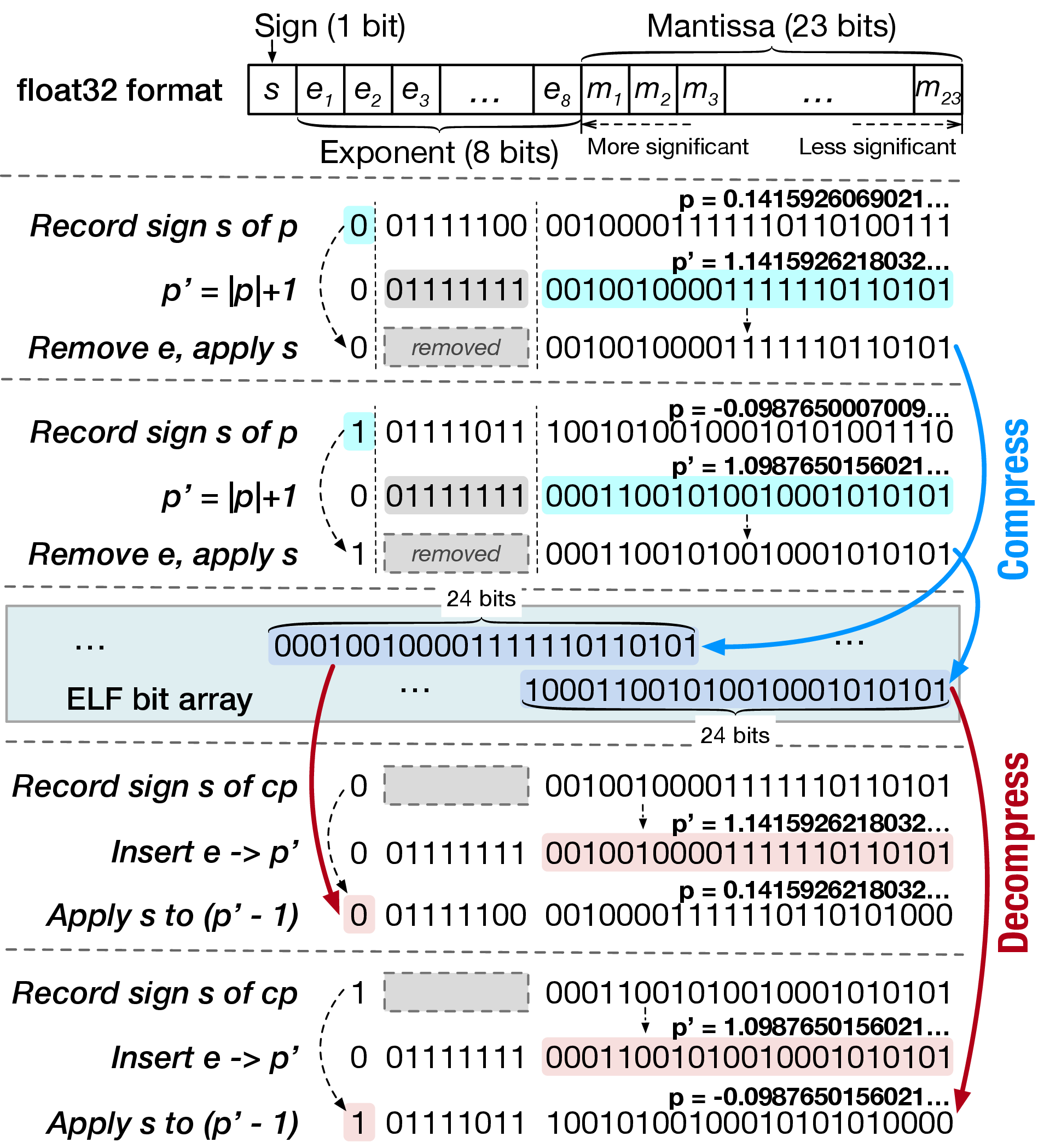}
\vspace{-10pt}
\caption{IEEE 754 Standard {\small\texttt{float32}} format and examples of the {\alg} compression process marked using \textcolor{blue}{blue arrows} and the decompression process marked using \textcolor{red}{red arrows}.   
\textit{\textmd{Take parameter $p=0.1415926069021\dots$ (binary machine representation) as an example, the compression process follows: 
transform $p$ to the intermediate value $p^{\prime}=1.1415926218032\dots$; take the sign bit of $p$ and the 23-bit mantissa of $p^{\prime}$ to construct the 24-bit compressed parameter $cp$, and finally, append it to {\alg}'s binary bit array. The decompression process is the inverse of the compression process. 
}}
}
\label{fig:elf_example}
\end{center}
\vspace{-15pt}
\end{figure}

Two key observations motivate the design of {\alg}.
(1)~Recall we have observed in \cref{sec:size_content_analysis} that around {$99\%$} parameters in our model dataset fall within the range of $(-1, 1)$.
(2)~Take the single-precision floating-point ({\small\texttt{float32}}) format as an example: In accordance with IEEE 754 Standard~\cite{ieee_std754}, a {\small\texttt{float32}} value $p$ is stored with 32 binary bits, where 1 bit is for the sign $s$, 8 bits for the exponent $\vec{e} = \langle e_1, e_2, e_3, \dots, e_8 \rangle$, and 23 bits for the mantissa $\vec{m} = \langle m_1, m_2, m_3, \dots, m_{23} \rangle$ {(there is a default bit, {\small\texttt{0b}}1, which is hidden on the most significant side of the 23-bit mantissa)}, as shown in Figure~\ref{fig:elf_example} (top).  
$p$'s value satisfies:
\vspace{-5pt}
\begin{equation} \label{eq:ieee754}
\begin{aligned}
    p = (-1)^s \times 2^{e-127} \times (1.m_1m_2\dots m_{23})_2 \\
      = (-1)^s \times 2^{e-127} \times (1+\sum_{i=1}^{23} m_i \times 2^{-i})
\end{aligned}  
\vspace{-5pt}
\end{equation}

\noindent{where} $e$ is the decimal value of $\vec{e}^2$.
Equation~\ref{eq:ieee754} decides that, 
for all $p$ where $p \in [1,2)$, the binary representation of $p$ has the same exponent bits {\small\texttt{0b}}$01111111$ with a decimal representation of 127. 
{This is because, when $s$ is 0 and $e$ is {\small\texttt{0b}}$01111111$ (which equals 127 in decimal), the value of the exponent field of $e-127 = 127 - 127 = 0$. This means that this {\small\texttt{float32}} parameter is positive and its value  
 is $(-1)^0 \times 1 \times (1.m_1m_2\dots m_{23})_2 \in [1,2)$.} 

{\alg} is based on these two observations. The main idea of {\alg} is to map all floating-point  
parameters $p$ where $p \in (-1,1)$ to $p^{\prime}$ where $p^{\prime} \in [1,2)$, so that the common exponent field {\small\texttt{0b}}$01111111$, in case of {\small\texttt{float32}}, can be eliminated in order to save storage cost. Figure~\ref{fig:elf_example} illustrates {\alg}'s compression and decompression using simple {\small\texttt{float32}} examples. 

\begin{figure*}[t]
\begin{center}
\vspace{-5pt}
\includegraphics[width=0.85\textwidth]{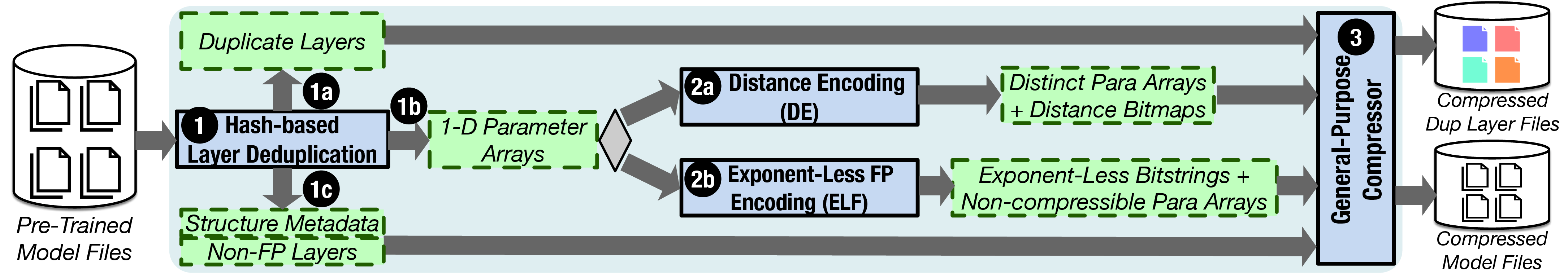}
\vspace{-12pt}
\caption{The {\wf} workflow.  
\textit{\textmd{Boxes with solid lines represent the stages of {\wf}, and boxes with dashed lines denote intermediate data. 
}}
}
\label{fig:workflow}
\end{center}
\vspace{-12pt}
\end{figure*}

\noindent\textbf{Compression.}
A sequential version of {\alg} scans model parameters linearly, and for each floating-point parameter $p$ where $p \in (-1, 1)$, the {\alg} compression performs the following three steps:
\begin{enumerate}
\item Record the sign bit $s$ of $p$ for later use in Step (3).
\item Convert $p$ where $p \in (-1,1)$ to $p^{\prime}$ where $p^{\prime} \in [1,2)$. 
\item Remove the exponent bits $\vec{e}$ of $p^{\prime}$, concatenate the 23-bit $\vec{m}$ of $p^{\prime}$ after the recorded sign bit $s$, and append the 24-bit compressed parameter $cp$ to the end of the bit array file. 
\end{enumerate}

\noindent\textbf{Decompression.}
The decompression process of a sequential version of {\alg} 
reads the compressed bit array file and restores all 24-bit compressed parameters sequentially. To restore a 24-bit unit $cp$ to $p$, 
the {\alg} decompression performs the following three steps:
\begin{enumerate}
\item Record the first bit $s$ (sign bit)  of $cp$ for later use in Step (3), and take the next 23 bits as the mantissa to restore an intermediate representation $p^{\prime}$.
\item Set the sign bit of $p^{\prime}$ with {\small\texttt{0b}}0 and insert the exponent bits {\small\texttt{0b}}$01111111$ between the sign bit and the 23-bit mantissa $\vec{m}$ of $p^{\prime}$ to restore $p^{\prime}$ so that $p^{\prime} \in [1,2)$. 
\item Construct a new intermediary $p^{\prime \prime}$ where $p^{\prime \prime} = p^{\prime} - 1 \in [0,1)$, and apply the recorded sign bit $s$ to $p^{\prime \prime}$ to restore the original parameter $p$ where $p \in (-1, 1)$. 

\end{enumerate}

\noindent{\textbf{Generality and Storage Savings.} 
{\alg} can be applied to all three types of floating-point values: {\small\texttt{float32}}, {\small\texttt{float16}}, and {\small\texttt{float64}}, although the storage efficiency varies depending on the data type. The storage savings of {\alg} come entirely from the removal of $e$ from the binary representation of each floating-point model parameter. For {\small\texttt{float32}} data, {\alg} can yield a $25\%$ ($\frac{8}{32}$) reduction in storage space. For {\small\texttt{float16}} and {\small\texttt{float64}}, the storage savings are $31.25\%$ ($\frac{5}{16}$) and $17.19\%$ ($\frac{11}{64}$), respectively. According to Table~\ref{tbl:layer_type_table}, over $99\%$ of parameters of our PTM dataset are composed of {\small\texttt{float32}} or {\small\texttt{float16}}, making these datasets particularly well-suited for {\alg}'s utility.}

\noindent\textbf{Parallelizability and Performance.}
{{\alg} is easily parallelizable as it is embarrassingly parallel via \emph{data parallelism}: a PTM dataset can be divided into chunks and each chunk can be compressed independently using a thread or a CPU core. Similarly, {\alg}'s decompression process can be easily parallelized using data parallel as well. This property guarantees {\alg}'s superior compression and decompression speed, which we evaluate in \cref{subsec:eval_throughput}.} 

\noindent{\textbf{Compression Loss.}} Since {\alg} involves data transformation and encoding for all floating-point parameters that fall within $(-1,1)$, this transforming process introduces bounded errors and the error bound varies depending on the data type.  
Before giving {\alg}'s error bound, we briefly discuss the process of floating-point addition. 
The exponents of the two numbers are compared to determine the larger exponent. The number with the smaller exponent is then shifted right by this difference in exponents so that both numbers have the same exponent. 
Then the mantissas of the two numbers (after shifting) are aligned by aligning the decimal points and added together using binary addition.
Lastly, the result is normalized and rounded to ensure the result fits within the specified precision. 

In {\alg}, when a {\small\texttt{float32}} $p \in (-1, 1)$ transforms to $p^{\prime} \in [1,2)$, there is no information loss when obtaining $p$'s absolute value $|p|$. Then for $1+|p|$, the exponent $e$ of $|p|$ needs to be shifted right (by adding the difference of $127-e$) so that the exponents of $|p|$ and 1 equal. This operation results in a right shift of the mantissa of $|p|$. After aligning and adding the mantissas of $|p|$ and 1, the result is normalized and rounded, and this process is where the error occurs. 
In other words, only the information from $|p| \in [0,1)$ captured by the first 23 bits of its mantissa is retained in $p^{\prime} \in [1,2)$, after that the less significant bits are rounded and discarded. 
Therefore, the maximum error introduced by {\alg} for {\small\texttt{float32}} parameters is $5.96046448 \times 10^{-8}$, or $2^{-24}$. Similarly, the error bound is $4.8828125 \times 10^{-4}$ or $2^{-11}$ for {\small\texttt{float16}} and $1.110223 \times 10^{-16}$ or $2^{-53}$ for {\small\texttt{float64}}, respectively. 

\noindent{\textbf{Storage Overhead.}
{\alg} stores out-of-range, non-compressible parameters $p \notin (-1,1)$ separately and uses an external table to keep track of the positions of these parameters. This extra storage overhead might outweigh the storage reduction obtained from {\alg}'s transformation and encoding, especially 
if the percentage of these out-of-range parameters is considerable. We address this problem using a hybrid approach, {\wf}, which will be described in \cref{subsec:workflow}.}

\vspace{-4pt}
\subsection{The {\wf} Compression Framework}
\label{subsec:workflow}

We present the design of our {offline} compression framework {\wf}. {\wf} incorporates the insights of {\alg} and a series of data reduction methods that we have explored in \cref{sec:compressibility_analysis}.
{Potential use case of {\wf} is to run {\wf} as a background process to scan all the PTM datasets that have already been written to storage~\cite{practical_dedup_fast11, dedup_survey_ieee} and select the most effective methods for data reduction.}

Figure~\ref{fig:workflow} depicts the stages of {\wf}. 
In Stage \circled{1}, {\wf} performs a scan over the entire PTM file dataset to compute the fingerprint of each model layer (\cref{subsec:dedup}). The fingerprint is computed based on the content of the layer by using a cryptographic hash function and is stored in a table that maps the layer ID to its fingerprint. When a new layer is encountered, its fingerprint is computed and compared with the fingerprints of existing layers in the table. If a matching fingerprint is found in the table, {\wf} detects that this layer is a duplicate, and instead of storing the layer again, a reference to the existing layer (stored as a separate layer file) is recorded. If no match is found, the new layer is unique, and it is stored along with its fingerprint. By the end of this stage, {\wf} stores all duplicate layers exactly once (intermediate output \circled{\footnotesize{1a}} in Figure~\ref{fig:workflow}) and continues to compress the rest of non-duplicate layers (intermediate output \circled{\footnotesize{1b}}) in next stages. 

{Non-duplicate layers are flattened to 1-dimensional (1-D) arrays of parameters of different floating-point types}, 
which are streamed to our DE (distance-encoding) compressor (see \cref{subsec:parameter} for the detailed description of the DE compression) in Stage \circled{\footnotesize{2a}} and {\alg} for exponent-less floating-point encoding (\cref{subsec:elf}) in Stage \circled{\footnotesize{2b}} for parameter-level compression. 
{{\wf} applies both DE and {\alg} to intermediate data \circled{\footnotesize{1b}} and chooses the compressor that provides a higher CR. The rationale is that, for models that have a substantial proportion of parameters $p \notin (-1,1)$, {\alg} might not be beneficial as it needs to keep track of these non-compressible parameters, which introduces extra storage overhead. 
Thus, for these models, 
{{\wf} opts to use DE over {\alg}, or the other way around, depending on which is more effective}. 
{\wf} then deletes the intermediate files generated by the less effective compressor. 
} 
In Stage \circled{3}, the outputs of DE or {\alg}, together with the duplicate layers (the intermediate output \circled{\footnotesize{1a}}), 
model structure metadata files, and non-floating-point model layers  (the intermediate output \circled{\footnotesize{1c}}) are further compressed by a general-purpose lossless compressor  Zstandard (zstd)~\cite{zstd}.
 
{The decompression process for {\wf} is the inverse of the compression process. 
(1)~Compressed files are decompressed by zstd to restore model structure metadata, non-floating-point layer files, duplicate layers, and DE-compressed / {\alg}-compressed intermediate files. 
(2)~DE-compressed / {\alg}-compressed intermediate files are decompressed by {\wf} 
to obtain the 1-D parameter arrays.  
(3)~Layers of the original models are recovered based on the model structure metadata. 
For each layer, there are three possibilities: 
($i$)~a duplicate layer will be retrieved from the corresponding decompressed duplicate layer file referenced by its fingerprint; ($ii$)~a floating-point layer will be restored from the 1-D array, based on the model structure specified by the model structure metadata; ($iii$)~a non-floating-point layer will be restored from the non-floating-point layer file. Upon completing these steps, the model dataset is decompressed.}

%% file: evaluation.tex
\vspace{-5pt}
\section{Evaluation}
\label{sec:eval}
\vspace{-2pt}

\noindent\textbf{Setup and Dataset.}
{We performed all of our tests on a server with 56 Intel(R) Xeon(R) Gold 6330 CPU cores and 256 GB memory running Ubuntu 20.04 with a kernel version of 5.4.0. Our evaluation is focused on our sampled dataset of {900} real-world pre-trained ML models (\cref{sec:datasets}) collected from Hugging Face. The total size of the binary format of the models that we tested is {575.88~GB}.}

\noindent\textbf{Baselines and Configurations.} 
We selected a total of {11} representative compressors divided into {four} categories. 

\begin{itemize}[noitemsep,leftmargin=*]
\vspace{-2pt}
  \item General-purpose lossless compressors: Gzip~\cite{gzip} and zstd~\cite{zstd}. 
  Both are based on LZ77~\cite{lz_tit77}. 
  They operate at the binary byte level and look for repetitive patterns among the bytes. We tested zstd's compression levels from 3 (default) to 19 (highest compression ratio). We found that all configurations led to the same compression ratio, but level 19 had extremely slow compression speed. Thus, we chose to use a compression level of 3 for zstd. 
  
  \item {Time series and data lake compressors:
  Sprintz~\cite{blalock2018sprintz}, Buff~\cite{liu2021decomposed}, Chimp~\cite{chimp_vldb22} and Gorilla~\cite{gorrila_vldb15} for TS datasets, 
  and {\btr}~\cite{btrblocks_sigmod2023} for data lakes. 
  Sprintz uses a lookup table to predict each value based on preceding entries and encodes the delta between predicted and original values for better compression. 
  Buff divides the sign, exponent, and mantissa, and tailors the storage scheme based on the specific bounds and precision requirements of the dataset.
  Chimp and Gorilla exploit the TS predictability, XOR successive values with previous ones, and compress away redundant zeros.
  {\btr} uses Pseudodecimal Encoding to convert float64 into two integers, significant digits with the sign and the exponent, to save storage.
  We used the default settings that their GitHub repositories specified for Sprintz, Chimp, and Gorilla, and matched the delta precision of Buff with the error bound of {\alg}. For {\btr} we used the single-column configuration given by its examples.}
    
  \item Error-bounded, lossy compressors for floating-point, scientific datasets: SZ3~\cite{sz3_icde21} and zfp~\cite{zfp}. SZ3 is a modular, error-bounded lossy compression framework for scientific datasets. SZ3 uses Lorenzo predictor~\cite{lorenzo_predictor} and regression predictor~\cite{lossy_bigdata18} to predict next parameters.
  SZ3 relies on the quantizer~\cite{lossy_hpdc20} to enable error control for prediction. zfp is designed for multidimensional numerical datasets. zfp uses transform to reduce the dynamic range of the floating-point data and then quantizes the transformed data. {Since integer-typed and boolean-typed layers in PTMs might be involved in mission-critical functionality, such as input layers, where no information loss is allowed, we tested SZ3 and zfp on floating-point layers of all models only.} The error bound for SZ3 and zfp were set to match the error bound of {\alg} (see the compression loss paragraph in \cref{subsec:elf}).
  
  \item {Model pruning and quantization methods: 
  global magnitude pruning (Global MP)~\cite{gupta2022complexity}, and Gaussian and outliers uniform quantization (GOUQ) based on GOBO~\cite{zadeh2020gobo}.  
  Global MP prunes parameters based on their magnitude while preserving the original model structure. Its parameter error 
  can be controlled by a threshold. 
  GOUQ is modified from GOBO, one of the cutting-edge quantization methods for speeding up NLP inference. GOUQ categorizes parameters into a ``Gaussian'' (\emph{G}) and an ``Outliers'' (\emph{O}) group. Parameters that fall into the \emph{O} group are stored in their original format (e.g., float32), while parameters of the \emph{G} group are quantized to a small set of representative values for space saving. GOUQ reduces the errors introduced by GOBO by increasing the number of representatives in the \emph{G} group, which introduces a tradeoff in storage saving and model accuracy. 
  We set the error bound of these two methods to align with the maximum error potentially introduced by {\alg}.} 
  
\vspace{-2pt}
\end{itemize}

\begin{figure*}[t]
\vspace{-12pt}
\begin{center}
\begin{minipage}{\textwidth}
\begin{minipage}[b]{0.305\textwidth}
\includegraphics[width=1\textwidth]{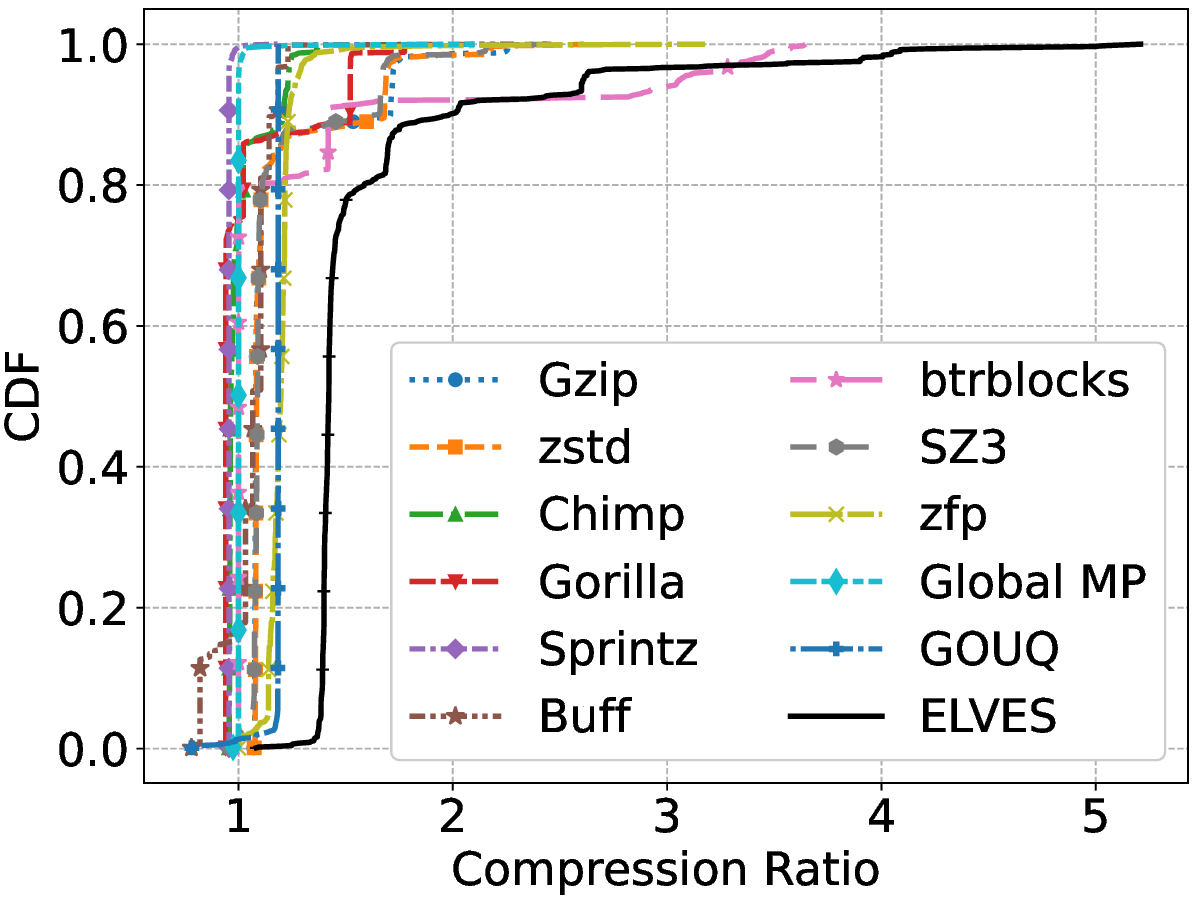}
\vspace{-15pt}
\caption{{Compression ratio comparison of different compressors for the 600-model dataset.
\textit{\textmd{Each data point in a curve is the CR of a model.
}}
}
}
\label{fig:compression_compressors}
\end{minipage}
\hspace{8pt}
\begin{minipage}[b]{0.305\textwidth}
\includegraphics[width=1\textwidth]{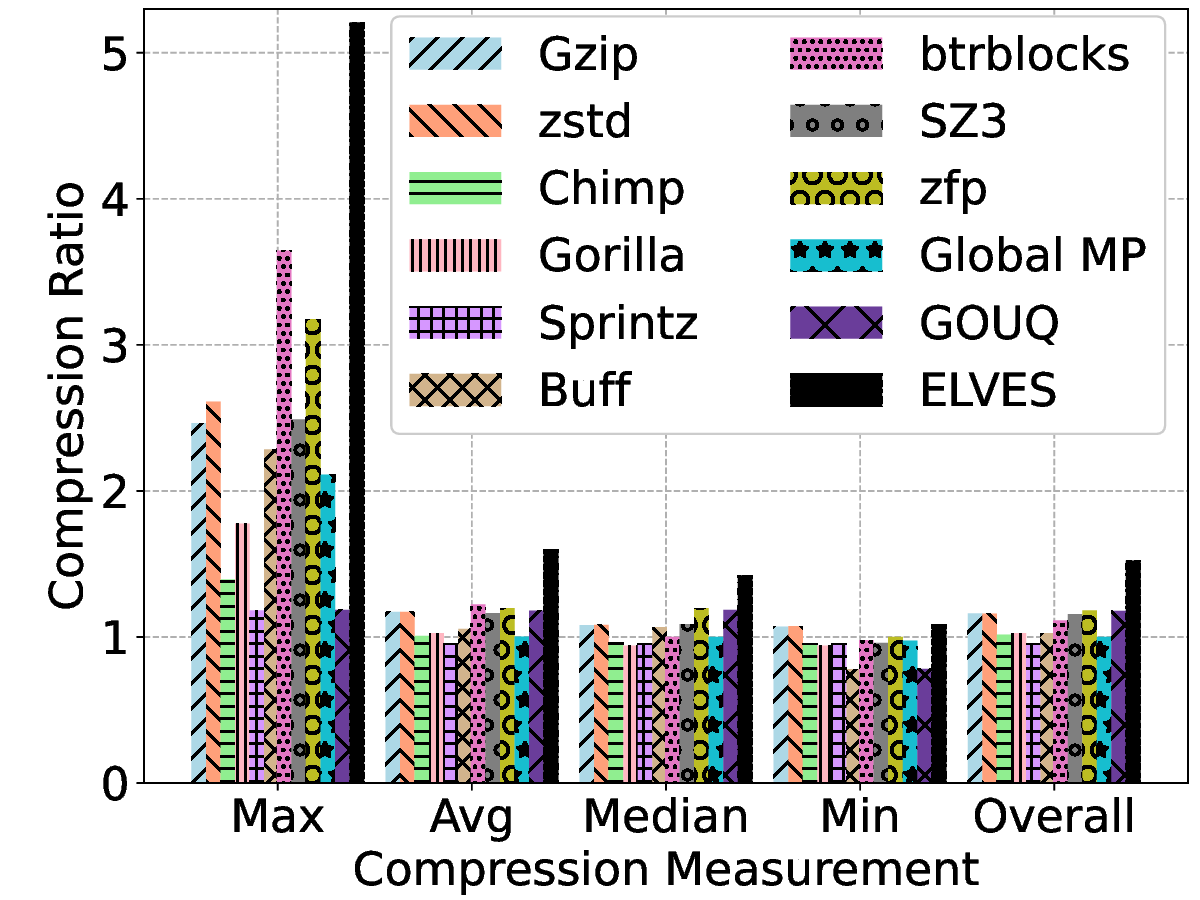}
\vspace{-15pt}
\caption{{Compression ratio breakdown. 
\textit{\textmd{Overall CR: the aggregate size of the original, uncompressed dataset divided by that of the compressed dataset.}}
}
}
\label{fig:compression_measurement}
\end{minipage}
\hspace{6pt}
\begin{minipage}[b]{0.35\textwidth}
\includegraphics[width=\linewidth]{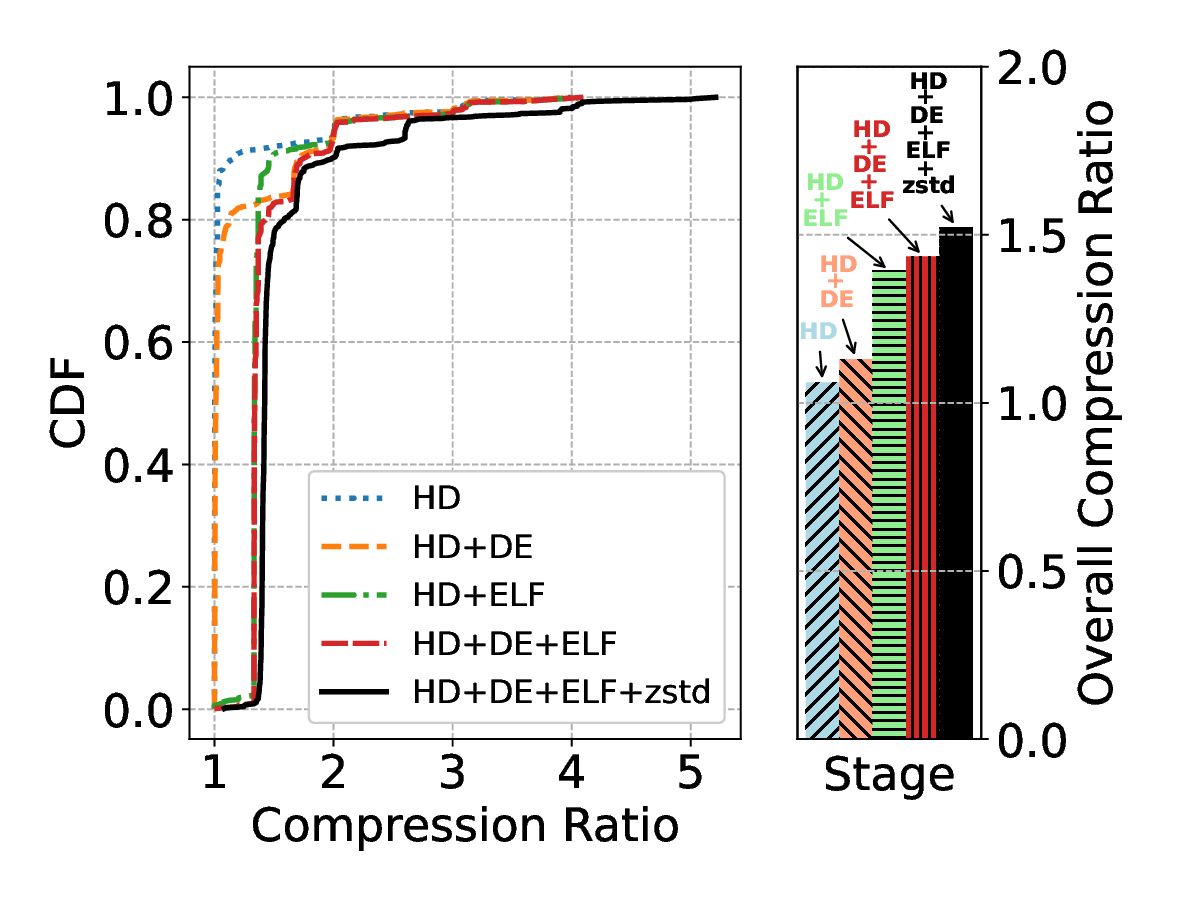}
\vspace{-22pt}
\caption{{Compression ratio CDF (left) and the cumulative fraction of overall compression ratio (right) for different {\wf} setups.}}
\vspace{10pt}
\label{fig:compression_modules}
\end{minipage}
\end{minipage}
\end{center}
\vspace{-15pt}
\end{figure*}

\vspace{-2pt}
\noindent\textbf{Goals.} Our evaluation aims to answer the following questions: 
\vspace{-2pt}
\begin{itemize}[noitemsep,leftmargin=*]
\vspace{-5pt}
  \item How does {\wf} compare to other baseline compressors in terms of compression ratio (\cref{subsec:eval_comparison})?  
   
  \item How does each stage of {\wf} contribute to {storage saving} 
  (\cref{subsec:eval_workflow})?  

  \item What is the compression and decompression speed of {\alg} (\cref{subsec:eval_throughput})?
  
  \item What is the impact of the lossiness of error-bounded {\alg} on the model accuracy (\cref{subsec:eval_acc})?
\vspace{-2pt}
\end{itemize}

\vspace{-4pt}
\subsection{Comparison with Baselines}
\label{subsec:eval_comparison}

First, we compare the {11} {baseline methods} with {\wf}. Figure~\ref{fig:compression_compressors} and \ref{fig:compression_measurement} show the compression ratio (CR)\footnote{CR is defined as the ratio between the size of the original, uncompressed dataset and the size of the compressed dataset. The higher, the better.} 
statistics. 

\noindent\textbf{General-Purpose Compressors.}
As Figure~\ref{fig:compression_compressors} shows, Gzip and zstd perform almost the same, {with an overall CR of $1.16\times$}. {zstd} outperforms {Gzip} slightly on maximum CR (see Figure~\ref{fig:compression_measurement}).
Both of them are able to reduce the model file sizes for almost all models, with around {$72\%$} of models having file sizes reduced by {$10\%$} (a CR of {$1.1\times$}).
The modest data reduction is due to that general-purpose compressors like Gzip and zstd are not specifically optimized for compressing floating-point datasets. 
However, it is noteworthy that for about {$11\%$} of models, Gzip and zstd still achieve a good CR of $1.5\times$. This is because these models have nearly $100\%$ parameter-level duplication, a {pattern} that can be exploited by {dictionary coding}.   

\noindent\textbf{State-of-the-Art, Encoding-based, Floating-Point Compressors.}
{TS data compressors perform poorly on our PTM dataset. Sprintz, Chimp, and Gorilla yield a CR of \emph{less than one} for $98.54\%$, $74.77\%$, and $75.23\%$ of models, respectively. This result indicates that they actually increased the file sizes of most models after compression. Approximately $82\%$ of models can save an average of $5\%$ storage space with Buff, but poor CR is achieved for about $11\%$ of the models with a minimum CR of $0.78\times$. The ineffectiveness of the above TS compressors is due to the significant differences in the data patterns found in TS datasets and PTM datasets. In TS data, two notable characteristics are often observed: 
(1)~the precision of the data may be relatively low, 
and (2)~there is typically a predictable trend or correlation among adjacent values. Sprintz employs a prediction scheme with a lookup table to record and encode the difference between predictions and real values into more compact forms. And Buff decomposes bounded low-precision floats into more compressible components, and encodes them with shorter representation than the original. Both Chimp and Gorilla leverage the correlation pattern of successive values by applying XOR operations to them and eliminating the resulting XOR'ed zeros to save storage costs. In PTM datasets, however, these features no longer exist, and adjacent parameters are randomly different. In terms of the overall CR, Sprintz, Buff, Chimp, and Gorilla achieve $0.96\times$, $1.03\times$, $1.02\times$, and $1.03\times$ for the entire dataset, respectively.}

{The columnar compressor, 
{\btr}, yields an overall CR of $1.12$. 
Specifically, it exhibits a CR of precisely 1.0 for $74.67\%$ of the models, indicating its ineffectiveness for PTM datasets. Furthermore, {\btr} attains a CR of $1.1$ or less for $80.11\%$ of the models; and a mere $18.56\%$ of PTMs achieving a CR of $1.3$ or higher, with the maximum CR at $3.65\times$.}
The reason is that the Pseudodecimal Encoding algorithm is more effective on the floating points with fixed, low precision---take 3.25 for example, {\btr} encodes $3.25$ as $(+325, 2)$, where $325 \times 10^{-2}$. 
However, most if not all parameters of PTMs exhibit a higher degree of precision (e.g., 1.0389173), rendering the encoding scheme of {\btr} ineffective. 

\noindent\textbf{State-of-the-Art, Error-Bound, Lossy Compressors.}
SZ3 achieves a CR of {$1.1\times$} for {$76.78\%$}
of models, which is {slightly higher} than that of Gzip and zstd, since SZ3 is a prediction-based compressor that is specifically designed and optimized for floating-point-based scientific datasets, which exhibit relatively ``smooth'' data patterns. But for PTMs that contain mostly trend-cluttered parameters, SZ3 requires a larger overhead to store points that fall outside the range that can be encoded through quantization and predictions. 
{zfp generally performs better than the other baselines}, with a CR ranging from $1.1\times$ to $1.3\times$ for {$93.78\%$} of models. {But it produces a CR greater than $1.3\times$ on only $3\%$ of the models.}
zfp achieves an overall CR of $1.18\times$ for the entire dataset, which is marginally elevated than that of zstd due to a lack of parameter correlations in model datasets. 

\noindent{\textbf{Model Pruning and Quantization Methods.} 
The storage savings achieved by Global MP on the PTM dataset are negligible. A significant majority, $98.11\%$, of PTMs exhibit a CR between $1.0\times$ and $1.05\times$, with an overall CR of $1.02\times$, suggesting the limited efficacy of Global MP in this context. To ensure the model integrity, Global MP preserves the original model structure by exploiting a bit string consisting of 0 and 1 indicating whether a parameter is removed or retained, which introduces a storage overhead. 
Furthermore, to prevent significant accuracy loss from pruning, even without re-training or fine-tuning, the magnitude threshold must be kept relatively low, which means only a minor portion of the model parameters are eligible for pruning, leading to modest compression results. 
The quantization method, GOUQ, slightly outperforms Global MP, achieving a CR ranging from $1.15\times$ to $1.19\times$ for $97.51\%$ of the models, with an overall CR of $1.18\times$. However, $1.22\%$ of the models exhibit a CR below 1, attributed to a considerable proportion of parameters being classified as outliers; and the overhead caused by the bit table distinguishing parameters as either Gaussian or outliers, surpasses the storage savings achieved from parameter mapping to representative values. Similarly, to ensure the accuracy of quantized models, parameter errors introduced during the quantization are strictly controlled.} 

\noindent\textbf{Comparing with {\wf}.}
Figure~\ref{fig:compression_compressors} shows a big margin between {\wf} and the baselines, because {\wf}' hybrid approach is designed based on the data characteristics of PTM datasets. {Specifically, {\wf} achieves $28.81\%$, $47.57\%$, and $28.81\%$ higher overall CR than zfp (the best in the error-bounded, lossy compressor set), Gorilla (the best in the TS data compressor set) and GOUQ (the best in the model quantization and pruning set)}, respectively, as shown in Figure~\ref{fig:compression_measurement}.

\vspace{-10pt}
\subsection{Evaluating {\wf} Stages}
\label{subsec:eval_workflow}

One way to understand the differences among hash-based layer deduplication (HD), distance-encoding (DE) compression, and {\alg} is to view the effectiveness of {\wf}' individual stages in terms of CR.
The CR distribution and cumulative improvement
of overall CR are depicted in  Figure~\ref{fig:compression_modules}. 

\noindent\textbf{{HD (Hash-based Deduplication).}} 
In this test we only enabled Stage \circled{1} of {\wf} (Figure~\ref{fig:workflow}).
While for most models the storage savings from HD are quite limited due to a low duplication ratio and small sizes of duplicate layers, around {$5.33\%$} of models exhibit a substantial level {(CR $\geq 2$)} of data reduction, with a CR of up to {$5.21\times$}. 
We anticipate that as the number of PTMs in real-world production environments continues to increase (Figure~\ref{fig:trend_plot}), the proportion of duplicate layers among models will also grow. This, in turn, will likely amplify the effectiveness of HD. 

\noindent\textbf{{HD + DE (Distance-Encoding).}}
In this test we enabled Stage \circled{1} and \circled{\footnotesize{2a}} of {\wf}. Figure~\ref{fig:compression_modules} (left) shows that DE results in a CR of $1.5\times$ and more for about {$11\%$ of more} models due to the fact that these models have close to $100\%$ of their parameters duplicated. Enabling DE atop HD improves the overall CR for all models by {$6.55\%$} (the right subfigure of Figure~\ref{fig:compression_modules}).

\noindent\textbf{{HD + {\alg} (Exponent-Less Floating-Point Encoding).}}
{Then, we enabled Stage \circled{1} HD and \circled{\footnotesize{2b}} {\alg}. 
{\alg} is generally effective for floating-point models.
{\alg} has a significant CR boost: \emph{for over $87\%$ of models where HD yields close-to-zero CRs, enabling {\alg} achieves an average CR of $1.35\times$, and improves the overall CR by $31.32\%$.} 
}

\noindent\textbf{{HD + DE + {\alg}.}} 
{Next, we combined Stage \circled{1}, \circled{\footnotesize{2a}} and \circled{\footnotesize{2b}} together. 
The compression strategies used by DE and {\alg} capitalize on different data patterns of the PTMs, rendering them complementary for some PTMs. As shown in the left subfigure of Figure~\ref{fig:compression_modules}, {\alg} guarantees an effective CR for a broad range of PTMs, while 
DE achieves better CRs on a small set of selected models. Integrating the three approaches yields enhancements in compression efficiency, surpassing HD+DE and HD+{\alg} by $27.06\%$ and $3.11\%$ respectively, as shown in the right subfigure of Figure~\ref{fig:compression_modules}.
}

\noindent\textbf{{HD + DE + {\alg} + zstd (Zstandard).}} 
Finally, by enabling Stage \circled{3} zstd to compress all intermediate data generated by previous stages, we observe an additional CR improvement of {$8.69\%$}. 

In summary, our ablation test demonstrates that: 
(1)~{\alg} has the greatest impact on reducing dataset sizes compared to other techniques, accounting for {$65.46\%$} of the end-to-end, overall CR improvement enabled by all stages. 
(2)~{\wf}' hybrid design is aware of the diverse data patterns of PTMs.
{\wf} effectively tailors the best compression method when compressing models with different patterns, yielding an overall CR of {$1.52\times$} for the whole dataset. {Moreover, $99\%$ of models (891 of 900) see a CR of over $1.35\times$, and $24.44\%$ of them achieve a CR $\ge 1.5\times$ with the highest CR of $5.21\times$}.

\vspace{-4pt}
\subsection{Evaluating {\alg} Performance}
\label{subsec:eval_throughput}

\begin{figure}[t]
\begin{center}
\begin{minipage}{\textwidth}

\begin{minipage}[b]{0.23\textwidth}
\includegraphics[width=1\textwidth]{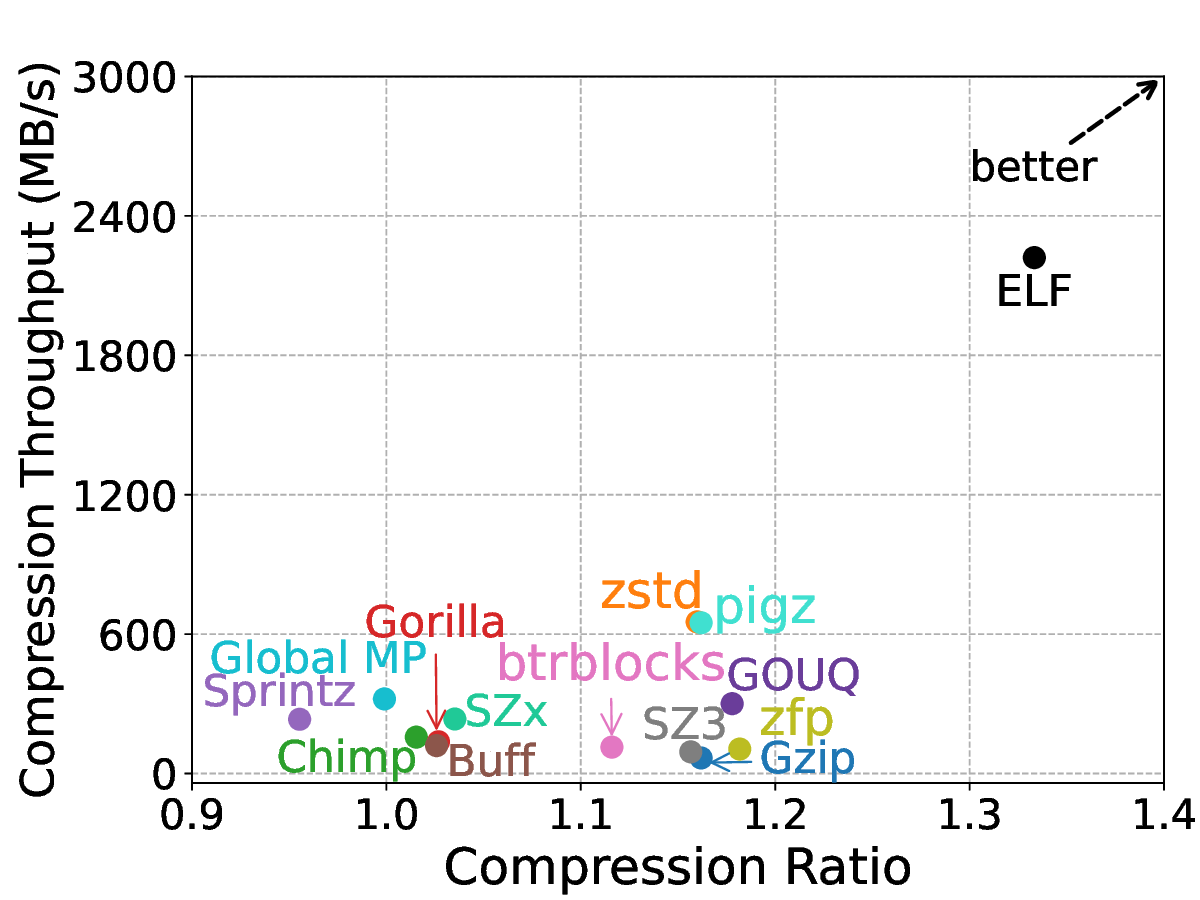}
\vspace{-20pt}
\caption{{CR vs. compression throughput. 
}
}
\label{fig:compression_throughput}
\end{minipage}
\hspace{2pt}
\begin{minipage}[b]{0.23\textwidth} 
\includegraphics[width=1\textwidth]{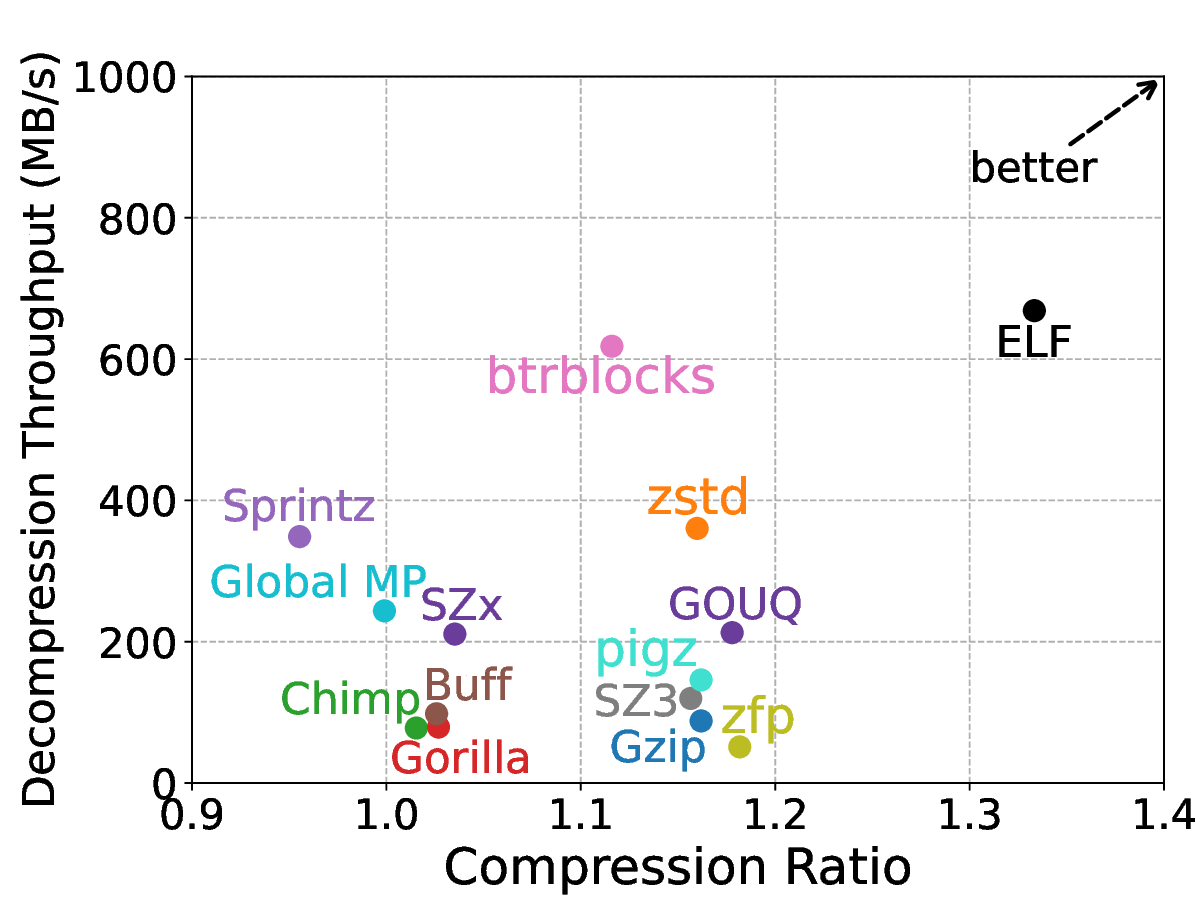}
\vspace{-20pt}
\caption{{CR vs. decompression throughput. 
}
}
\label{fig:decompression_throughput}
\end{minipage}

\end{minipage}
\end{center}
\vspace{-25pt} 
\end{figure}

{We have implemented the core compression and decompression algorithm of {\alg} using C++ and pthread. Next, we compare the compression and decompression speed of {\alg} with {13} baselines. To ensure the best throughput performance, we enabled multi-threading configuration for those baselines with data parallelism support: zstd, pigz (parallel Gzip)~\cite{pigz},  {\btr}, and SZx~\cite{szx_hpdc22}.}

{Figure~\ref{fig:compression_throughput} and \ref{fig:decompression_throughput} show the results. The throughput was calculated using the ratio of the aggregate size of the whole dataset and the total compression (or decompression) time. Note that, since we target storage compression and decompression, our throughput metric includes the I/O time each method took for reading files from the disk and writing files to the disk. We used a 1.6TB Intel Optane DC P5800X SSD, which provides a sequential read (write) throughput of {4.2 GB/s (938 MB/s)}. \emph{This is to ensure that the compression and decompression processes are not bottlenecked by the disk I/O.} From Figure~\ref{fig:compression_throughput} and \ref{fig:decompression_throughput} we can see that {\alg} significantly outperforms all other baselines in compression throughput, and also has the fastest decompression speed among all ten compressors. By using all the available 56 CPUs, {\alg} achieves an average compression (decompression) throughput of {2,170.56~MB/s (653.58~MB/s)}, respectively. zstd with multi-thread setting using all 56 CPUs is still $3.6\times$ ($1.89\times$) slower in compression (decompression) speed, compared to {\alg}. While {\btr} exhibits a comparable decompression throughput with {\alg}, {\alg} achieves a compression throughput that is {$35.29\times$} greater than {\btr}'s. With a single thread, {\alg} achieves a compression and decompression throughput of 121.68~MB/s and 135.02~MB/s, respectively, still competitive compared to baselines such as Gzip, {Buff, zfp}, Chimp, and Gorrila. 
The strong performance results indicate that {\alg} can serve as a practically useful tool for PTM storage compression.}

\vspace{-4pt}
\subsection{Quantifying Impact on Model Accuracy} 
\label{subsec:eval_acc}

This section evaluates the impact of {\alg} on model accuracy using two methods: fuzz-testing-inspired validation  (\cref{subsec:fuzz_acc}) and benchmark validation (\cref{subsec:benchmark_acc}). 

\subsubsection{Fuzz-Testing-Inspired Validation}\label{subsec:fuzz_acc}
{\alg} introduces bounded errors for floating-point parameters during compression.
{We evaluated the impact of {\alg} on model accuracy using a dataset comprising 300 out of 900 models from 9 different tasks, ranging from image classification to text generation, across three model categories.}

{
We quantify the impact of compression on model accuracy using a metric called \emph{accuracy degradation} (AD) for a given task and dataset, which is defined as follows:
\vspace{-4pt}
\begin{equation}\label{eq:accuracy_degradation} 
    \text{AD} = 
    \frac{\sum_{i=1}^{N} {\Delta A_i}}{N}=
    \frac{\sum_{i=1}^{N} \sum_{j=1}^{M} {compare(O_{i,j} , O'_{i,j})}}{N\cdot M}
\vspace{-6pt}
\end{equation} \\
where $\Delta A_i$ represents the difference between the original model and the decompressed model for the same task and under the same dataset. $N$ is the total number of models tested this time. $O$ is the output of the original model and $O^{'}$ is the output of the model after decompression.
Each model, original or decompressed, generates an output tensor that consists of $M$ float numbers. We use $M$ to define the size of the output and use a function $compare()$ to compare the outputs of two models under a defined output precision. In the case of the fuzz-testing validation, we compare each bit of two floats. Here is how we make the comparison: for a tensor with an output length of 1,000, we compare the float numbers at the corresponding positions of the two tensors. If the floats are consistent within all digits, we consider the two model outputs to be the same, and  $compare()$ outputs 0; otherwise, it outputs 1. If 999 out of 1,000 floats are the same, then we obtain an AD of 0.1\%. 
}

\noindent\textbf{Testing Methodology.} 
Our methodology for evaluating the accuracy degradation of the model is inspired by \emph{fuzz testing}~\cite{fuzz_unix_cacm90, miller1995fuzz, windowsnt_usenixwindows00}. In fuzz testing, random inputs are generated and then fed to a program to verify the correctness of the program. Here, we generate random inputs for each model and compare the output generated by the original and compressed model on these inputs. {We used AD to evaluate all the models except the text-generation task since this kind of model has search algorithms based on randomness to generate non-repeated text for the same prompt fed to the model multiple times.} 
The non-deterministic nature of the model's output makes it challenging to evaluate and compare the accuracy, as the output of both decompressed and original models may likely differ (in subtle ways) for the same input. 
{Thus, we modified our $compare()$ to adapt this situation. Now, $compare()$ will calculate the cosine similarity of the embeddings and attention weights generated by the model, and if the cosine similarity of the embeddings is \emph{exactly} 1, it outputs 0 else 1.
Cosine similarity for output \{$\mathbf{O_i}$, $\mathbf{O_j}$\} from two models is defined as: $C_{sim}(O_i, O_j) = \frac{\mathbf{O_i} \cdot \mathbf{O_j}}{\|\mathbf{O_i}\| \|\mathbf{O_j}\|}$. 
}

The input for NLP-based models is generated from selected Unicode character ranges that include characters from various languages. This is done to evaluate the models not trained in Latin-based languages. We also did the same for speech recognition, as some of the models are trained in many different languages. If an NLP model does not support a non-Latin-based Unicode sequence, then the input is changed to a sequence of Latin-based characters. Text-to-text (T2T) generation is a task where a text is fed to a model that outputs a text. The difference between text generation and T2T generation is that a text generation model behaves like a chatbot where, for a given prompt, the model outputs the information it is trained on. While T2T generation works as a multi-modal architecture that can perform various tasks like question answering, translation and summarization, etc. To fuzz test an image classification model, we generated an image based on random noise, fed it to the network, and compared the top $k$ labels predicted by the original model. 
Some models for classification tasks and fill-mask predict top $k$ labels for a given input. We want to compare all the labels predicted by each label, and for these types of tasks, {we define our $compare()$ function as let $Y^k$ be the set of top $k$ labels predicted by model $m$. We compare the output of these models and outputs 0 if and only if $Y^k_o - Y^k_{o'} = \emptyset$ where $Y^k_o, Y^k_c$ are top $k$ labels that are output by two models.}

\begin{table}[t]
\centering
\caption{{Accuracy degradation of different model task categories. 
\textit{\textmd{$\%$ in Column 2 represent the proportion of all tested models.}}}
}
\vspace{-8pt}

\scalebox{0.85}{
\begin{tabular}{lrr}
\hline 
{\bf Model Task (Category)} & {\bf Count (\%)} & {\bf {Accuracy Degradation}}  \\
\hline 
Image Classification (CV)              &   {69 (23.00\%)}       & {0.87\%} \\
Text Generation (NLP)                  &   {68 (22.67\%)}       & {0\%} \\
Text Classification (NLP)              &   {60 (20.00\%)}       & {0\%} \\
Token Classification (NLP)             &   {30 (10.00\%)}       & {0\%} \\
Translation (NLP)                      &   {25 (8.33\%)}        & {0.4\%} \\
Question Answering (NLP)               &   {24 (8.00\%)}        & {0\%} \\
Audio Classification (Audio)           &   {9 (3.00\%)}         & {0\%} \\
Summarization (NLP)                    &   {9 (3.00\%)}         & {1.11\%} \\
Speech Recognition (Audio)             &   {6 (2.00\%)}         & {0\%} \\
{\bf Overall}                          &   {\bf {300 (100\%)}}  & {\bf {0.27\%}} \\
\hline
\end{tabular}
}

\label{tbl:category_and_accuracy_new}
\vspace{-10pt}

\end{table}

Some of the models hosted on Hugging Face were missing information about the tokenizer they used when training the model. To circumvent this issue, we manually set the tokenizer of the model to either XLM or BERT. XLM is used for the models that are trained on languages other than English as it has a huge vocabulary dictionary to tokenize various tokens from different languages. 

\noindent\textbf{Validation Results.}
{Even with a rigorous definition to compare the outputs of decompressed models and their original versions, we were able to achieve an AD of $0\%$ ($0\%$ indicates that every decompressed model for the given task generates the \emph{exact same output} as its original version) for 6 out of the 9 tasks and $< 1.2\%$ for remaining tasks, leading to an overall accuracy degradation of $0.27\%$ (Table~\ref{tbl:category_and_accuracy_new}).}

\begin{table}[ht]
\centering
\caption{{{Accuracy degradation of all loss compression frameworks with the benchmark datasets.}
\textit{\textmd{Benchmark validation has been evaluated on a total of 46 models across 9 tasks in 4 domains. The error bound of {\wf} is denoted with $e$. All tested baselines and their abbreviations are as follows: SZ3, zfp, Global MP (mp), Global MP with $2\times$ error bound (mp2e), GOUQ (gouq), GOUQ with $2\times$ error bound (gouq2e), and half-precision quantization (half). The overall AD is averaged across all tasks for a particular compression method, while the overall CR is calculated by dividing the combined size of the original models by the total size of the compressed models. 
}}}}
\vspace{-8pt}
\scalebox{0.44}{
\begin{tabular}{|c|c|c|cccccccc|} 
\hline
\multirow{2}{*}{\textbf{Domain}} &  \multirow{2}{*}{\textbf{Task(\# of tested model)}} & \multirow{2}{*}{\textbf{Dataset}} & \multicolumn{8}{c|}{\textbf{Accuracy Degradation}} \\ \cline{4-11}
 & & & \multicolumn{1}{c}{\textbf{\wf}} & \multicolumn{1}{c}{\textbf{SZ3}} & \multicolumn{1}{c}{\textbf{zfp}} & \multicolumn{1}{c}{\textbf{mp}} & \multicolumn{1}{c}{\textbf{mp2e}} & \multicolumn{1}{c}{\textbf{gouq}} & \multicolumn{1}{c}{\textbf{gouq2e}} & \multicolumn{1}{c|}{\textbf{half}} 
 \\
\hline
\multirow{6}{*}{\textbf{CV}}
& \multicolumn{1}{c|}{\multirow{2}{*}{\textbf{image classification(4)}}} 
&\textbf{mini\_imagenet} 
&{0.2\%} &{0.3\%} &{0.2\%} &{0.1\%} &{0.2\%} &{0.4\%} &{1.1\%} &{65.0\%} \\
& &\textbf{cifar100} 
&{0.2\%} &{0.3\%} &{0.1\%} &{0.2\%} &{0.2\%} &{0.4\%} &{1.2\%} &{48.4\%} \\

& \multicolumn{1}{c|}{\multirow{2}{*}{\textbf{object detection(4)}}} 
&\textbf{detection-datasets/coco} 
&{0.1\%} &{0.2\%} &{0.2\%} &{0.1\%} &{0.2\%} &{0.2\%} &{0.2\%} &{1.6\%} \\
& &\textbf{cppe-5} 
&{0.2\%} &{0.3\%} &{0.2\%} &{0.2\%} &{0.3\%} &{0.2\%} &{0.3\%} &{2.6\%} \\

& \multicolumn{1}{c|}{\multirow{2}{*}{\textbf{image segmentation(6)}}} 
&\textbf{scene\_parse\_150} 
&{0.2\%} &{0.6\%} &{0.4\%} &{0.1\%} &{0.2\%} &{0.2\%} &{0.8\%} &{38.6\%} \\
& &\textbf{sidewalk-semantic} 
&{0.3\%} &{1.4\%} &{0.5\%} &{0.2\%} &{0.3\%} &{0.2\%} &{0.7\%} &{35.1\%} \\ \hline

\multirow{4}{*}{\textbf{Multimodal}}

& \multicolumn{1}{c|}{\multirow{2}{*}{\textbf{feature extraction(7)}}} 
&\textbf{Open-Orca/OpenOrca} 
&{0.1\%} &{0.2\%} &{0.1\%} &{0.1\%} &{0.1\%} &{0.2\%} &{0.3\%} &{18.1\%} \\
& &\textbf{imdb-movie-reviews} 
&{0.1\%} &{0.1\%} &{0.1\%} &{0.1\%} &{0.1\%} &{0.2\%} &{0.5\%} &{24.5\%} \\

& \multicolumn{1}{c|}{\multirow{2}{*}{\textbf{image-to-text(4)}}} 
& \textbf{conceptual\_captions} 
&{0\%} &{0\%} &{0\%} &{0\%} &{0\%} &{0\%} &{0\%} &{0\%} \\
& &\textbf{red\_caps} 
&{0\%} &{0\%} &{0\%} &{0\%} &{0\%} &{0\%} &{0\%} &{0\%} \\
\hline

\multirow{2}{*}{\textbf{Audio}}
& \multicolumn{1}{c|}{\multirow{2}{*}{\textbf{speech recognition(5)}}}
&\textbf{librispeech\_asr\_dummy} 
&{0\%} &{0\%} &{0\%} &{0\%} &{0\%} &{0\%} &{0\%} &{0\%} \\
& &\textbf{lj\_speech} 
&{0\%} &{0\%} &{0\%} &{0\%} &{0\%} &{0\%} &{0\%} &{0\%} \\ \hline

\multirow{6}{*}{\textbf{NLP}}
& \multicolumn{1}{c|}{\multirow{2}{*}{\textbf{sentiment classification(7)}}}
& \textbf{glue-sst2} 
&{0\%} &{0\%} &{0\%} &{0\%} &{0\%} &{0\%} &{0\%} &{0\%} \\
& &\textbf{imdb} 
&{0\%} &{0\%} &{0\%} &{0\%} &{0\%} &{0\%} &{0\%} &{0\%} \\

& \multicolumn{1}{c|}{\multirow{2}{*}{\textbf{sentence similarity(5)}}}
& \textbf{glue-stsb} 
&{0\%} &{0\%} &{0\%} &{0\%} &{0.1\%} &{0.1\%} &{0.2\%} &{3.6\%} \\ 
& &\textbf{paws-x} 
&{0\%} &{0\%} &{0\%} &{0\%} &{0.1\%} &{0.1\%} &{0.2\%} &{4.2\%} \\ 

& \multicolumn{1}{c|}{\multirow{2}{*}{\textbf{Fill-mask(4)}}}
& \textbf{wikitext} 
&{0\%} &{0\%} &{0\%} &{0\%} &{0.1\%} &{0.1\%} &{0.1\%} &{0.1\%} \\
& &\textbf{ptb\_text\_only} 
&{0\%} &{0\%} &{0\%} &{0\%} &{0.1\%} &{0.1\%} &{0.1\%} &{0.1\%} \\ \hline

\multicolumn{3}{|c|}{\textbf{Overall AD}} 

&{\textbf{0.07\%}} &{\textbf{0.18\%}} &{\textbf{0.1\%}} &{\textbf{0.06\%}} &{\textbf{0.22\%}} &{\textbf{0.13\%}} &{\textbf{0.32\%}} &{\textbf{13.44\%}}  \\

\multicolumn{3}{|c|}{\textbf{(Overall CR)}} 

&{\textbf{(1.52)}} &{\textbf{(1.16)}} &{\textbf{(1.18)}} &{\textbf{(1.00)}} &{\textbf{(1.01)}} &{\textbf{(1.18)}} &{\textbf{(1.20)}} &{\textbf{(1.99)}}  \\
\hline
\end{tabular}
}
\vspace{-10pt}
\label{tbl:benchmark}
\end{table}
\subsubsection{Benchmark Validation}\label{subsec:benchmark_acc}

While our large-scale fuzz-testing based accuracy validation in \cref{subsec:fuzz_acc} covers a broad set of 300 models, standard benchmark datasets offer a more comprehensive way of validating the model accuracy. 
{We conducted benchmark tests on a total of 9 tasks across 4 domains. Table~\ref{tbl:benchmark} shows the results. We see that {\wf} is the only method that achieves both an overall accuracy degradation close to zero (0.07\%) and a high CR. This result demonstrates that {\wf} 
has negligible influence on the performance of models and all outputs generated using decompressed models are almost identical to those generated by the original models.}

%% file: conclusion.tex
\section{Conclusion}
\label{sec:conclusion}

This paper dissects the data characteristics of real-world pre-trained ML model datasets and studies their compressibility along different dimensions. Our analysis considers different representative data reduction and compression techniques and spans three data granularities, including model layers, model chunks, and model parameters. Through our comprehensive analysis, we find that PTM dataset compression is challenging, and that existing data reduction and compression techniques are generally ineffective for reducing the storage size of PTM datasets. 
Based on the observations, we have proposed {\alg}, a simple and effective, error-bounded, lossy floating-point compression algorithm and developed {\wf}, a compression framework that integrates {\alg} and several other techniques. {\wf} achieves an overall compression ratio of $1.45\times$, which is up to $1.3\times$ higher than state-of-the-art lossy floating-point compressors, while introducing close to zero model accuracy loss. We hope that our study will provide valuable insights into the design, implementation, and optimization of data reduction techniques and systems for efficient storage of PTM datasets.